\documentclass[aps,jmp,preprint,groupedaddress]{revtex4-1}
	\usepackage{amssymb,latexsym}
	
	\usepackage{amsmath}
	\usepackage{color}
	\newcommand\eq[1] {(\ref{#1})}

	\newcommand{\bfm}[1]{\mbox{\boldmath ${#1}$}}
	\newcommand{\nonum}{\nonumber \\}
	\newcommand{\beqa}{\begin{eqnarray}}
	\newcommand{\eeqa}[1]{\label{#1}\end{eqnarray}}
	\newcommand{\beq}{\begin{equation}}
	\newcommand{\eeq}[1]{\label{#1}\end{equation}}
	\newcommand{\bpm}{\begin{pmatrix}}
	\newcommand{\epm}{\end{pmatrix}}

	\newcommand{\Grad}{\nabla}
	\newcommand{\Div}{\nabla \cdot}
	\newcommand{\Curl}{\nabla \times}
	
	\newcommand{\Real}{\mathop{\rm Re}\nolimits}
	\newcommand{\Imag}{\mathop{\rm Im}\nolimits}

	\newcommand{\Md}{\partial}

	\newcommand{\Ga}{\alpha}
	\newcommand{\Gb}{\beta}
	\newcommand{\Gd}{\delta}
	\newcommand{\Ge}{\epsilon}

	\newcommand{\Gg}{\gamma}

	\newcommand{\Gk}{\kappa}
	
	\newcommand{\Gl}{\lambda}
	
	\newcommand{\Gm}{\mu}
	\newcommand{\Gv}{\nu}
	
	\newcommand{\Gt}{\theta}
	
	\newcommand{\Gr}{\rho}
	
	\newcommand{\Gs}{\sigma}

	\newcommand{\Go}{\omega}
	
	\newcommand{\Gy}{\psi}

	\newcommand{\GL}{\Lambda}

	\newcommand{\GO}{\Omega}

	\newcommand{\BGb}{\bfm\beta}
	
	\newcommand{\BGe}{\bfm\epsilon}
	\newcommand{\BGve}{\bfm\varepsilon}

	\newcommand{\BGk}{\bfm\kappa}

	\newcommand{\BGm}{\bfm\mu}

	\newcommand{\BGr}{\bfm\rho}
	
	\newcommand{\BGs}{\bfm\sigma}

	\newcommand{\BGy}{\bfm\psi}

	\newcommand{\BGF}{\bfm\Phi}
	\newcommand{\BGG}{\bfm\Gamma}
	\newcommand{\BGL}{\bfm\Lambda}

	\newcommand{\CE}{{\cal E}}

	\newcommand{\CH}{{\cal H}}
	
	\newcommand{\CJ}{{\cal J}}

	\newcommand{\CR}{{\cal R}}
	
	\newcommand{\CT}{{\cal T}}

	\newcommand{\CZ}{{\cal Z}}
	\newcommand{\BCC}{{\bfm{\cal C}}}
	\newcommand{\BCD}{{\bfm{\cal D}}}

	\def\ii{{\rm i}}

	\def\Bb{{\bf b}}

	\def\Be{{\bf e}}
	\def\Bf{{\bf f}}
	
	\def\Bh{{\bf h}}
	
	\def\Bj{{\bf j}}
	\def\Bk{{\bf k}}

	\def\Bn{{\bf n}}

	\def\Bq{{\bf q}}
	\def\Br{{\bf r}}
	\def\Bs{{\bf s}}
	
	\def\Bu{{\bf u}}
	\def\Bv{{\bf v}}
	\def\Bw{{\bf w}}
	\def\Bx{{\bf x}}
	\def\By{{\bf y}}
	\def\Bz{{\bf z}}
	\def\BA{{\bf A}}
	\def\BB{{\bf B}}
	\def\BC{{\bf C}}
	\def\BD{{\bf D}}
	\def\BE{{\bf E}}
	
	\def\BG{{\bf G}}
	
	\def\BI{{\bf I}}
	\def\BJ{{\bf J}}
	
	\def\BL{{\bf L}}
	\def\BM{{\bf M}}

	\def\BP{{\bf P}}
	\def\BQ{{\bf Q}}
	\def\BR{{\bf R}}
	\def\BS{{\bf S}}
	\def\BT{{\bf T}}
	
	\def\BV{{\bf V}}
	\def\BW{{\bf W}}
	
	\def\BY{{\bf Y}}
	
	\def\B0{{\bf 0}}

	\def \RR {{\mathbb R}}
	
	\def \ba {\begin{array}}
	\def \ea {\end{array}}
	\newtheorem {Thm} {Theorem} [section]

	\newtheorem {Adef} [Thm] {Definition}
	
	\newtheorem {Arem} [Thm] {Remark}
	
	\newtheorem {Aexa} [Thm] {Example}
	
	\newtheorem {Anot} [Thm] {Notation}

	\def \refe #1.{(\ref{#1})}
	\def \reff #1.{figure~\ref{#1}}
	\def \refs #1.{section~\ref{#1}}
	\def \refss #1.{subsection~\ref{#1}}
	\def \refD #1.{Definition~\ref{#1}}
	\def \refT #1.{Theorem~\ref{#1}}
	\def \refL #1.{Lemma~\ref{#1}}
	\def \refC #1.{Corollary~\ref{#1}}
	\def \refP #1.{Proposition~\ref{#1}}
	\def \refR #1.{Remark~\ref{#1}}
	\def \refE #1.{Example~\ref{#1}}
	\def \refN #1.{Notation~\ref{#1}}
\newcommand{\aj}[1]{\textcolor{black}{#1}}

\begin{document}
	%%%%%%%%%%%%%%%%%%%%%%%%%%%%%%%%%%%%%%%%%%%%%%%%%%%%%%%%%%%%%%%%%%
	%%%%%%%TITLE %%%TITLE %%%TITLE %%%TITLE %%%TITLE %%%TITLE %%%%%%%%
	%%%%%%%%%%%%%%%%%%%%%%%%%%%%%%%%%%%%%%%%%%%%%%%%%%%%%%%%%%%%%%%%%%
	
	\title{A new route to finding bounds on the generalized spectrum of many physical operators}
	\author{Graeme W. Milton}
        \email[]{milton@math.utah.edu}
        \affiliation{Department of Mathematics, University of Utah}	
	%%%%%%%%%%%%%%%%%%%%%%%%%%%%%%%%%%%%%%%%%%%%%%%%%%%%%%%%%%%%%%%%%%
	%%%%%%%BEGIN DOCUMENT %%%%BEGIN DOCUMENT %%%%BEGIN DOCUMENT%%%%%%%
	%%%%%%%%%%%%%%%%%%%%%%%%%%%%%%%%%%%%%%%%%%%%%%%%%%%%%%%%%%%%%%%%%%
        \date{\today}

	%%%%%%%%%%%%%%%%%%%%%%%%%%%
	%%%%%%%%ABSTRACT%%%%%%%%%%%
	%%%%%%%%%%%%%%%%%%%%%%%%%%%
	
	\begin{abstract}
          Here we obtain bounds on the generalized spectrum of that operator whose inverse, when it exists, gives the Green's function. We
          consider the wide of physical problems that can be cast in a form where a constitutive equation $\BJ(\Bx)=\BL(\Bx)\BE(\Bx)-\Bh(\Bx)$ 
          with a source term $\Bh(\Bx)$ holds for all $\Bx$ in some domain $\GO$, and relates fields $\BE$ and $\BJ$ that satisfy appropriate differential
          constraints, symbolized by $\BE\in\CE_\GO^0$ and $\BJ\in\overline{\CJ}_\GO$ where $\CE_\GO^0$ and $\overline{\CJ}_\GO$ are orthogonal spaces that span
          the space $\CH_\GO$ of square-integrable fields in which $\Bh$ lies. Boundedness and coercivity conditions 
          on the moduli $\BL(\Bx)$ ensure there exists a unique $\BE$ for any given $\Bh$, i.e. $\BE=\BG_\GO\Bh$,
          which then establishes the existence of the Green's function $\BG_\GO$. We show that the coercivity condition is
          guaranteed to hold if weaker conditions, involving generalized quasiconvex functions, are satisfied. The advantage is that
          these weaker conditions are easier to verify, and for multiphase materials they can be independent of the geometry of the phases. 
          For $\BL(\Bx)$ depending linearly on a vector of parameters
          $\Bz=(z_1, z_2,\ldots, z_n)$, we obtain constraints on $\Bz$ that ensure the Green's function exists, and hence which
          provide bounds on the generalized spectrum.  
	\end{abstract}
	
	%%%%%%%%%%%%%%%%%%%%%%%%%%%
	%%%%%%%%%PACS AND KEYWORDS%%%%%%%%%%
	%%%%%%%%%%%%%%%%%%%%%%%%%%%
	
\maketitle
	%%%%%%%%%%%%%%%%%%%%%%%%%%%%%%%%%%%%%%%%%%%%%%%%%%%%%%%%%%%%%%%%%%%%%%%%%%%%%%%%%%%%%%%%%%%%%%%%%%%%%%%%%%%%%%%%%%%%%%%%%%%%%%%%%%%%
	\section{Introduction}
%\setcounter{equation}{0}
%%%%%%%%%%%%%%%%%%%%%%%%%%%%%%%%%%%%%%%%%%%%%%%%%%%%%%%%%%%%%%%%%%%%%%%%%%%%%%%%%%%%%%
\aj{This paper continues the theme of the book ``Extending the Theory of Composites to Other Areas of Science'' \cite{Milton:2016:ETC},
(reviewed in \cite{Sharma:2017:BRE,Grabovsky:2018:BRE})
using tools in the theory of composites \cite{Cherkaev:2000:VMS,Torquato:2002:RHM,Milton:2002:TOC,Allaire:2002:SOH,Tartar:2009:GTH}
to obtain results pertinant to other problems. Specifically in this paper we use ideas from the Theory of Composites to
derive bounds on the generalized spectrum of the operator whose inverse, when it exists, gives the Green's function.} 
Here we modify the definition  of the Green's function to avoid unnecessary complexities of having to deal with complicated
function spaces, and to simplify the analysis. We believe this is a much better approach to operator theory than conventionally taken.
The approach builds upon that developed in \cite{Milton:2018:ERG} where the infinite body Green's function, defined in the appropriate way, 
was found to satisfy certain exact identities for wide classes of inhomogeneous media. \aj{These exact identities generalize the theory
of exact relations for composites \cite{Grabovsky:1998:EREa,Grabovsky:1998:EREb,Grabovsky:2000:ERE} (see also Chapter 17 of \cite{Milton:2002:TOC} and the book \cite{Grabovsky:2016:CMM}) that, like our analysis, derives from the splitting of the relevant Hilbert space into orthogonal subspaces.
Such exact relations identify tensor manifolds such that the effective tensor lies on the manifold whenever the local tensor takes values
in the manifold.}

There is, of course, an enormous body of literature establishing bounds on the spectra of operators, that reflects the importance of this problem. For compact Hermitian operators
on Hilbert spaces the Courant-Fischer-Weyl min-max principle gives a variational characterization of the eigenvalues, and can be used to bound the eigenvalues through the Cauchy
interlacing principle. Horn's theorem \cite{Horn:1954:DSM} bounds the eigenvalues of a Hermitian matrix in terms of its diagonal elements.
Various inequalities
\cite{Weyl:1912:AVE,Lidskii:1950:PVS,Wielandt:1955:OSE,Horn:1962:BPI,Thompson:1971:ESH,Helmke:1995:EIS,Klyachko:1954:DSM} relate the eigenvalues 
of a sum of Hermitian matrices to the eigenvalues of the matrices in the sum: an excellent summary can be found in the paper of Fulton \cite{Fulton:2000:EIF}.
Such eigenvalue problems can be regarded as determining when certain linear pencils of Hermitian matrices have a nontrivial nullspace, when one of the matrices is the identity matrix. 
Here we are concerned with such questions for linear operators rather than matrices (without the restriction that one of the operators is the identity operator)
and the bounds we obtain incorporate information about the underlying
partial differential equation, using Null-Lagrangians, quasiconvexity and its generalizations
\cite{Tonelli:1920:SCV,Terpstra:1938:DBF,Morrey:1952:QSM,Meyers:1965:ASA,Morrey:1966:MIC,Murat:1978:CPC,Tartar:1979:CCA,
Tartar:1979:ECH,Ball:1981:NLW,Murat:1981:CPC,Murat:1985:CVH,Tartar:1985:EFC,Benesova:2017:WLS}, 
notably using the notion of $Q^*$-convexity \cite{Milton:2013:SIG,Milton:2015:ATS}. Interestingly, there
are other connections between the spectrum of matrices, quasiconvexity, and rank-one convexity aside from those developed here:  see \cite{Buliga:208:FAM} and references therein.

\aj{It is well known that the classic equation of electrical conductivity in the absence of current sources $\Div\BGs\Grad V=0$ in which $V(\Bx)$ is the electric potential and $\BGs(\Bx)$ is the conductity tensor, can be alternatively 
formulated as 
\beq \Bj(\Bx)=\BGs(\Bx)\Be(\Bx),\quad\Div\Bj(\Bx)=0,\quad\Be(\Bx)=-\Grad V, \eeq{0.0}
where $\Bj$ and $\Be$ is the electrical current and field. This alternative formulation, involving fields satisfying differential constraints 
linked by a constitutive law involving a tensor that contains information about the material moduli, and its generalizations, forms the basis for much analysis in the theory
of composites (see, e.g. \cite{Milton:2002:TOC}), and has been advocated, for example, by Strang \cite{Strang:1988:FEE}. As shown in \cite{Milton:2009:MVP} and in Chapter 1 of \cite{Milton:2016:ETC}, and as will
be reviewed below (see also the appendix), the formulation can be extended to wave equations in the frequency domain
and to the Schr{\"o}dinger equation. It provides a route to obtaining results that may not be directly evident from standard formulations.}

\aj{The formulation naturally extends to the} large variety of linear equations in physics that can be written as a system of second-order linear partial differential equations:
\beq \sum_{i=1}^d\frac{\Md}{\Md x_i}\left(\sum_{j=1}^d\sum_{\Gb=1}^sL_{i\Ga j\Gb}(\Bx)\frac{\Md u_\Gb(\Bx)}{\Md x_j}\right)=f_\Ga(\Bx), \quad\Ga=1,2,\ldots,s,\eeq{0}
for the $s$-component potential $\Bu(\Bx)$. If the integral of $\Bf(\Bx)$ over $\RR^d$ is zero, these can be reexpressed as 
\beq J_{i\Ga}(\Bx)=\sum_{j=1}^d\sum_{\Gb=1}^sL_{i\Ga j\Gb}(\Bx)E_{j\Gb}(\Bx)-h_{i\Ga}(\Bx),\quad E_{j\Gb}(\Bx)=\frac{\Md u_\Gb(\Bx)}{\Md x_j},\quad \sum_{i=1}^d\frac{\Md J_{i\Ga}(\Bx)}{\Md x_i}=0,
\eeq{1}
where, counter to the usual convention, we find it convenient to let the divergence act on the first index of $\BJ$, and to let the gradient in $\BE=\Grad\Bu$ be associated
with the first index of $\BE$, and $\Bh(\Bx)$ is chosen so
\beq \sum_{i=1}^d\frac{\Md h_{i\Ga}(\Bx)}{\Md x_i}=f_\Ga(\Bx). \eeq{2}
Examples include, for instance, the electrostatics equation, linear elastostatics equation, piezoelectric equation, the quasistatic acoustic, Maxwell, and elastodynamic, equations 
(where the fields and moduli are generally complex): see Chapter 2 in \cite{Milton:2002:TOC} for numerous examples. 
If we restrict attention to the space of square integrable fields, so that $\BE(\Bx)$ and $\BJ(\Bx)$ are square-integrable, integration by parts shows that
\beq (\BJ,\BE)\equiv\int_{\RR^d} (\BJ(\Bx),\BE(\Bx))_\CT\,d\Bx=0,\quad (\BJ(\Bx),\BE(\Bx))_\CT=\sum_{i=1}^d\sum_{\Ga=1}^s J_{i\Ga}(\Bx)[E_{i\Ga}(\Bx)]^*,
\eeq{3}
where the asterisk denotes complex conjugation. Thus $\BE(\Bx)$ and $\BJ(\Bx)$ belong to orthogonal spaces: $\CE$ the set of square-integrable fields $\BE(\Bx)$ such that $\BE=\Grad\Bu$ 
for some $m$ component potential $\Bu$, and $\CJ$ the set of square-integrable fields $\BJ(\Bx)$ such that $\Div\BJ=0$. With these definitions, the equations \eq{1}
take the equivalent, more abstract, form 
\beq \BJ(\Bx)=\BL(\Bx)\BE(\Bx)-\Bh(\Bx),\quad \BJ\in\CJ,\quad \BE\in\CE,\quad \Bh\in\CH.
\eeq{4}
where $\CH=\CE\oplus\CJ$ consists of square integrable $d\times m$ matrix-valued fields. \aj{The benefit of this more abstract formulation is that it brings under the 
one umbrella a wide variety of physical equations, and enables one to develop results, such as exact identities satisfied by the Green's function \cite{Milton:2018:ERG}, that
may be obscured in more conventional approaches. Also the orthogonality of the spaces $\CE$ and $\CJ$ naturally leads to minimization variational principles (even 
for wave equations in lossy media) and the norms of relevant operators can often be easily estimated, thanks to the fact that the projection operators onto
$\CE$ and $\CJ$, like any projection, have norm one. Additionally, the formalism allows one to see that the ``Dirichlet-to-Neumann'' map that governs the response
of inhomogeneous bodies has Herglotz-type analytic properties not only as a function of frequency, but also as a function of the component moduli, and consequently there are
associated integral representation formulas for this map (see Chapters 3 and 4 in \cite{Milton:2016:ETC}, also arXiv:1512.05838 [math.AP]).}

Given any simply-connected region $\GO$ with smooth 
boundary $\Md\GO$, the fields in $\CE$ and $\CJ$ also satisfy the key identity
\beq  \int_\GO (\BJ(\Bx),\BE(\Bx))_\CT\,d\Bx=\int_{\Md\GO}B(\Md\BJ,\Md\BE)\,dS\text{   for any }\BE\in\CE,\quad\BJ\in\CJ,
\eeq{5}
where $\Md\BE$ and $\Md\BJ$ denote the boundary fields associated with $\BE$ and $\BJ$ respectively,
and $B(\Md\BJ,\Md\BE)$ is linear in $\Md\BJ$ and antilinear $\Md\BE$, i.e., for any $c_1,c_2\in\mathbb{C}$,
\beq B(c_1\Md\BJ,c_2\Md\BE)=c_1 c_2^*B(\Md\BJ,\Md\BE), \eeq{5.aa}
in which $c_2^*$ denotes the complex conjugate of $c_2$.
In the context of the equations \eq{1}, the boundary field $\Md\BE$ can be identified with the
potential $\Bu(\Bx)$ for $\Bx\in\Md\GO$, and $\Md\BJ$ can be identified with the boundary fluxes $\Bn(\Bx)\cdot\BJ(\Bx)$ where $\Bn(\Bx)$
is the outwards normal to $\Md\GO$. \aj{ We are avoiding a precise definition of $\Md\BJ$ and $\Md\BE$ as these boundary fields are not central to the paper,
they depend on the problem of interest, and are not so easily defined in
cases (such as plate equations) where the differential constraints on the fields are higher than first order. Roughly speaking,
they are associated with physical problems for which the response of a body is governed by some ``Dirichlet-to-Neumann'' map and $\Md\BJ$ and $\Md\BE$ are the boundary fields
associated with this map. Various examples of the boundary fields $\Md\BJ$ and $\Md\BE$ will be given below, and in the appendix.}

More generally, other equations of physics, including heat and wave equations, can be expressed in the form \eq{4} with an identity
like \eq{5} holding. For example, at fixed frequency $\Go$ with a $e^{-i\Go t}$ time dependence, as recognized in \cite{Milton:2009:MVP}
the acoustic equations, with $P(\Bx)$ the pressure, $\Bv(\Bx)$ the velocity, $\BGr(\Bx,\Go)$ the effective mass density matrix, and $\Gk(\Bx,\Go)$ the bulk modulus, take the form
\beq \underbrace{\begin{pmatrix}-i\Bv \\ -i\Div\Bv \end{pmatrix}}_{\BJ(\Bx)}
=\underbrace{\begin{pmatrix}-(\Go\BGr)^{-1} & 0 \\ 0 & \Go/\Gk\end{pmatrix}}_{\BL(\Bx)}\underbrace{\begin{pmatrix}\Grad P \\ P\end{pmatrix}}_{\BE(\Bx)},
\eeq{5.A}
(and $\Md\BE$ and $\Md\BJ$ can be identified with the boundary values of $P(\Bx)$ and $\Bn\cdot\Bv(\Bx)$ at $\Md\GO$, respectively).
Here we allow for effective mass density matrices that, at a given frequency, can be anisotropic and complex valued as may be the case in metamaterials 
\cite{Schoenberg:1983:PPS,Willis:1985:NID,Milton:2006:CEM,Milton:2007:MNS}.
Maxwell's equations, with $\Be(\Bx)$ the electric field, $\Bh(\Bx)$ the magnetizing field, $\BGm(\Bx,\Go)$ the magnetic permeability,
$\BGve(\Bx)$ the electric permittivity, take the form
\beq \underbrace{\begin{pmatrix}-i\Bh \cr i\Curl\Bh\end{pmatrix}}_{\BJ(\Bx)}
=\underbrace{\begin{pmatrix}-{[\Go\BGm]}^{-1} & 0 \\ 0 & \Go\BGve \end{pmatrix}}_{\BL(\Bx)}
\underbrace{\begin{pmatrix}\Curl\Be \\ \Be\end{pmatrix}}_{\BE(\Bx)},
\eeq{5.B}
(and $\Md\BE$ and $\Md\BJ$ can be identified with the tangential values of $\Be(\Bx)$ and $\Bh(\Bx)$ at $\Md\GO$, respectively).
The linear elastodynamic equations, with $\Bu(\Bx)$ the displacement, $\BGs(\Bx)$ the stress, $\BCC(\Bx,\Go)$ the elasticity tensor,
$\BGr(\Bx,\Go)$ the effective mass density matrix,  take the form
\beq \underbrace{\begin{pmatrix} -\BGs/\Go \\ -\Div\BGs/\Go\end{pmatrix}}_{\BJ(\Bx)} 
=\underbrace{\begin{pmatrix}-\BCC/\Go & 0 \\ 0 & \Go\BGr\end{pmatrix}}_{\BL(\Bx)}\underbrace{\begin{pmatrix}\Grad \Bu \\ \Bu\end{pmatrix}}_{\BE(\Bx)},
\eeq{5.C}
(and $\Md\BE$ and $\Md\BJ$ can be identified with the values of $\Bu(\Bx)$ and the traction $\Bn\cdot\BGs(\Bx)$ at $\Md\GO$, respectively)
where we note that it is not necessary to introduce the strain field (symmetrized gradient of the displacement $\Bu$) as the elasticity tensor
$\BCC$ annihilates the antisymmetric part of $\Grad\Bu$. The preceeding three equations have been written in this form so $\Imag\BL(\Bx)\geq 0$
when $\Imag\Go\geq 0$, where complex frequencies have the physical meaning of the solution increasing exponentially in time.
Using an approach of Gibiansky and Cherkaev \cite{Cherkaev:1994:VPC} this allows one to express the solution as the minimum of
some appropriately defined functional \cite{Milton:2009:MVP,Milton:2010:MVP}. The Schr{\"o}dinger equation for the wavefunction $\Gy(\Bx)$
of a single electron in a magnetic field, in the time domain, with $\Bb=\Curl\BGF$ the magnetic induction,
$V(\Bx,t)$ the time-independent electric potential, $e$ is the charge on the electron, and $m$ its mass, 
takes the form \cite{Milton:2016:ETC}:
\beq
\underbrace{\begin{pmatrix}
\Bq_x\\
q_t\\
\nabla\cdot\Bq_x+\frac{\partial q_t}{\partial t}
\end{pmatrix}}_{\BJ(\Bx)} 
=
\underbrace{\begin{pmatrix}
\frac{-\BI}{2m} & 0 & \frac{\ii e \BGF}{2m}\\
0 & 0 & -\frac{\ii}{2}\\
\frac{-\ii e \BGF}{2m} & +\frac{\ii}{2} & -e V
\end{pmatrix}}_{\BL(\Bx)}
\underbrace{\begin{pmatrix}
\nabla\Gy\\
\frac{\partial\Gy}{\partial t}\\
\Gy
\end{pmatrix}}_{\BE(\Bx)},
\eeq{5.D}
in which dimensions have been chosen so that $\hbar=1$, where $\hbar$ is Planck's constant divided by $2\pi$
(and $\Md\BE$ and $\Md\BJ$ can be identified with the values of $\Gy(\Bx)$ and the flux $\Bn_x\cdot\Bq_x(\Bx)+n_tq_t$ at $\Md\GO$, respectively,
in which $(\Bn_x,n_t)$ is the outwards normal to the region $\GO$ in space-time).
The multielectron Schr{\"o}dinger equation with a time dependence $e^{-iEt/\hbar}$ and with $\hbar=1$ takes the form \cite{Milton:2016:ETC}:
\beq
\underbrace{\begin{pmatrix}\Bq(\Bx) \\ \Div\Bq(\Bx)\end{pmatrix}}_{\BJ(\Bx)}
=\underbrace{\begin{pmatrix}-\BA & 0 \\ 0 & E-V(\Bx)\end{pmatrix}}_{\BL(\Bx)}\underbrace{\begin{pmatrix}\Grad \psi(\Bx) \\ \psi(\Bx)\end{pmatrix}}_{\BE(\Bx)},
\eeq{5.E}
where $V(\Bx)$ is the potential and $\BA$ in the simplest approximation is $\BI/(2m)$ in which $m$ is the
mass of the electron, but it may take other forms to take into account the reduced mass of the
electron, or mass polarization terms due to the motion of the atomic nuclei. Here $\Bx$ lies in a multidimensional space $\Bx=(\Bx_1,\Bx_2,\ldots,\Bx_N)$ where following,
for example, \cite{Parr:1994:DFT}, each $\Bx_i$ represents a pair $(\Br_i,s_i)$ where $\Br_i$ is a three dimensional vector
associated with the position of electron $i$ and
$s_i$ denotes its spin (taking discrete values $+1/2$ for spin up or $-1/2$ for spin down). Accordingly, $\nabla$ represents the operator
\beq
\nabla = (\nabla_1,\nabla_2,\ldots,\nabla_N),\quad\text{where}~~
\nabla_j = \left(\frac{\partial}{\partial
r_1^{(j)}},\frac{\partial}{\partial r_2^{(j)}},\frac{\partial}{\partial
r_3^{(j)}}\right).
\eeq{5.F}
When the energy $E$ is complex, $E=E'+iE"$, then minimization variational principles for the multielectron  Schr{\"o}dinger equation exist (see Chapter 13 in \cite{Milton:2016:ETC}).
\aj{Specifically, consider the functional 
\beq W(\Gy')=\sum_s\int_{\GO^N}[p(\Bx,\Gy')]^2+(E'')^2|\Gy'(\Bx)|^2~d\Br,\quad p(\Bx,\Gy')=\Div\BA\Grad\Gy'+(E'-V(\Bx))\Gy',
\eeq{5.Fa}
where the sum is over all $2^N$ spin configurations $s=(s_1,s_2,\ldots,s_N)$ as each $s_j$ takes values $+1/2$ or $-1/2$. Then in any body $\GO$ with appropriate boundary
conditions on $\Gy'$ on $\Md\GO^N$, $W(\Gy')$ is minimized when $\Gy'$ is the real part of the wave function $\Gy$ which satisfies the  Schr{\"o}dinger equation
\beq \Div\BA\Grad\Gy+(E-V(\Bx))\Gy=0, \eeq{5.Fb}
in which the potential $V(\Bx)$ is assumed to be real. Following the prescription outlined in \cite{Milton:2010:MVP}, a variety of different boundary conditions on $\Gy'$ can be handled by suitably adjusting the functional $W(\Gy')$: see Section 13.3 in \cite{Milton:2016:ETC}.}

With only pair potentials the multielectron Schr{\"o}dinger equation is also equivalent to the desymmetrized  multielectron Schr{\"o}dinger equation
which takes the form
\beq \BJ(\Bx)=\underbrace{\begin{pmatrix}-\BA & 0 \\ 0 & E-g(\Bx_1,\Bx_2)\end{pmatrix}}_{\BL(\Bx)}\underbrace{\begin{pmatrix}\Grad \psi(\Bx) \\ \psi(\Bx)\end{pmatrix}}_{\BE(\Bx)},
\quad \BGL\BJ(\Bx)=\begin{pmatrix}\Bq(\Bx) \\ \Div\Bq(\Bx)\end{pmatrix},\quad \BGL\BE(\Bx)=\BE(\Bx),
\eeq{5.G}
where $\BGL$ is an appropriate symmetrization operator defined in Chapter 12 of \cite{Milton:2016:ETC}. The desymmetrized  multielectron Schr{\"o}dinger equation has the advantage 
(with complex values of $E$ and source terms) that it can be solved iteratively by going back and forth between real and Fourier space using Fast Fourier
transforms where the Fourier transforms only need to be done on the two variables $\Bx_1$ and $\Bx_2$, not on all variables. Many additional equations
can be expressed in the canonical form: these include, at least, the thermoacoustic equations at constant frequency, and in the time domain 
(without assuming a $e^{-i\Go t}$ time dependence) the acoustic, Maxwell, elastodynamic, piezoelectric, and plate equations, 
Biot poroelastic equations, thermal conduction and diffusion equations,
thermoelastic equation, and the Dirac equation for the electron. The interested reader is referred to Chapter 1 of \cite{Milton:2016:ETC} for more details. 
The formalism also extends to scattering problems, where one needs to introduce an auxiliary space ``at infinity'' to ensure orthogonality 
of appropriately defined spaces $\CE$ and $\CJ$ \cite{Milton:2017:BCP}.
Of course, in all these equations one can allow for an additional source term $\Bh(\Bx)$, as is needed to define the Green's function. For wave equations in the
time domain, rather than the frequency domain, one has to impose boundary conditions appropriate to selecting the causal Green's function. To avoid such complications
our focus will be on static and quasistatic equations, and on wave equations in the frequency domain. \aj{Beyond the forms \eq{5.A}, \eq{5.B}, and \eq{5.C} for acoustics,
electromagnetism, and elastodynamics, canonical forms in the frequency domain for thermoacoustics, thermoelasticity, and the plate equations, are given in the appendix}.

The classical Poincare inequality seeks bounds on the lowest eigenvalue of the Dirichlet Laplacian operator that is associated with the equation
\beq  \underbrace{\begin{pmatrix}\BQ \\ \Div\BQ \end{pmatrix}}_{\BJ(\Bx)}
=  \underbrace{\begin{pmatrix} 1 & 0 \\ 0 & -z\end{pmatrix}}_{\BL}\underbrace{\begin{pmatrix}\Grad \Bu \\ \Bu\end{pmatrix}}_{\BE(\Bx)},
\eeq{5.Ga}
on a domain $\GO$ with $\Bu=0$ on $\Md\GO$. Bounds on higher order eigenvalues have been obtained too: see, for example, \cite{Ilyin:2006:LBS} and references
therein.

In general, the fields $\BJ(\Bx)$, $\BE(\Bx)$, and $\Bh(\Bx)$ take values in some tensor space $\CT$. The tensor space has the characteristic feature that
there is a inner product $(\cdot,\cdot)_\CT$ on $\CT$ such that for every rotation $\BR$ in $\mathbb{R}^d$ there is a corresponding operator $\CR(\BR)\CT\to\CT$ 
such that $\CR(\BI)=\BI$
\beq (\CR(\BR)\BA,\CR(\BR)\BB)_\CT=(\BA,\BB)_\CT\text{  for all }\BA,\BB\in\CT.
\eeq{5b}
This inner product assigned to $\CT$ is useful when one is considering a body containing a $m$-phase polycrystalline material with moduli
\beqa &~& \BL(\Bx)=[\CR(\BR(\Bx))]^\dagger\widetilde{\BL}(\Bx)\CR(\BR(\Bx)),\quad \widetilde{\BL}(\Bx)=\sum_{i=1}^m\chi_i(\Bx)\BL_i,\nonum
&~&  [\CR(\BR(\Bx))]^\dagger\CR(\BR(\Bx))=\BI,\quad \sum_{i=1}^m\chi_i(\Bx)=1,
\eeqa{10}
where the $\BL_i$ are the tensors of the individual phases, $\chi_i(\Bx)$ is the characteristic functions of phase $i$, taking the value $1$ in phase $i$ and zero
outside of it, and $\BR(\Bx)$ is a rotation field giving the orientation of the material at the point $\Bx$. A particular, but important, case is a multicomponent body
where $\BR(\Bx)=\BI$ and $\CR(\BR(\Bx))=\BI$ for all $\Bx$ so that \eq{10} reduces to
\beq \BL(\Bx)=\sum_{i=1}^m\chi_i(\Bx)\BL_i, \quad \sum_{i=1}^m\chi_i(\Bx)=1.
\eeq{10A}
We will not generally require
that the inner product assigned to $\CT$ satisfy the property \eq{5b}. There is good reason for removing this constraint. Indeed, in say three dimensions, with $m=3$ the potentials
$u_1(\Bx)$, $u_2(\Bx)$, and $u_3(\Bx)$ could represent three different scalar potentials such as voltage, temperature, and pressure. However it is mathematically equivalent
to the problem of linear elasticity where these represent components of the displacement field $\Bu(\Bx)$. Despite the mathematical equivalence, the potentials they behave quite differently
under rotations, and thus $\CR(\BR)$ is different. 

In most applications the spaces $\CH$, $\CE$ and $\CJ$ are real-symmetric, in the sense that
if a field belongs to them, then so does the complex conjugate field. Our analysis goes through without this assumption:
it will only be used just below equation \eq{F.3a} to simplify the criterion \eq{F.3}. 

\aj{Our focus is on the generalized spectrum rather than the spectrum, and there is good reason for that. The spectrum typically consists of the range of values of the
energy, frequency (or frequency squared), or wave number away from which the Green's function exists. However, in many applications the component moduli also
depend upon frequency, and this greatly complicates the picture. Thus it makes sense to separate the dependence on the component moduli, as
embodied in the generalized spectrum. Varying the frequency then corresponds to following a trajectory in the space of component moduli
that may intersect the generalized spectrum, and at frequencies
at these intersections the Green's function does not exist. } 

%%%%%%%%%%%%%%%%%%%%%%%%%%%%%%%%%%%%%%%%%%%%%%%%%%%%%%%%%%%%%%%%%%%%%%%%%%%%%%%%%%%%%%
\section{Showing the Green's function exists when suitable boundedness and coercivity conditions are satisfied}
%\setcounter{equation}{0}
%%%%%%%%%%%%%%%%%%%%%%%%%%%%%%%%%%%%%%%%%%%%%%%%%%%%%%%%%%%%%%%%%%%%%%%%%%%%%%%%%%%%%%

For fields $\BP(\Bx)$ and $\BP'(\Bx)$ defined within $\GO$ we define the inner product and norm
\beq (\BP,\BP')_\GO=\frac{1}{V(\GO)}\int_\GO (\BP(\Bx),\BP'(\Bx))_\CT\,d\Bx,\quad |\BP|_\GO=\sqrt{(\BP,\BP)_\GO},
\eeq{6}
where $V(\GO)$ is the volume of $\GO$. 
Define $\CH_{\GO}$ to consist of those fields $\BP$ taking values in $\CT$ that are square integrable over $\BQ$ in the sense that the norm $|\BP|_\GO$ is finite.
Define $\CE_\GO^0\subset \CH_{\GO}$ as the restriction to $\GO$ of those fields in $\CE$ satisfying the boundary constraint $\Md\BE=0$.
Alternatively $\CE_\GO^0$ can be viewed as consisting
of those fields $\BE(\Bx)$ in $\CH_{\GO}$ that when extended to $\CH$ by defining $\BE(\Bx)=0$ outside $\GO$, have the property that this extended field
lies in $\CE$. We define $\CJ_\GO\subset \CH_{\GO}$ as the restriction to $\GO$ of those fields in $\CJ$ 
(with no boundary constraint), and we let $\overline\CJ_\GO$ denote its closure. The subspaces $\CE_\GO^0$ and $\overline\CJ_\GO$ are orthogonal and span $\CH_{\GO}$
as proved in Lemma 2.3 of \cite{Grabovsky:2016:CMM}. 
To define the Green's operator, with the boundary condition $\Md\BE=0$,
the equations of interest now take the form
\beq  \BJ(\Bx)=\BL(\Bx)\BE(\Bx)-\Bh(\Bx),\quad \BJ\in\overline\CJ_\GO,\quad \BE\in\CE_\GO^0,\quad \Bh\in\CH_\GO.
\eeq{6a}
We assume that within $\GO$, $\BL(\Bx)$ takes the form of a pencil of linear operators:
 \beq \BL(\Bx)=\sum_{i=1}^nz_i\BL^{(i)}(\Bx).
\eeq{7}
For example, in the setting of the polycrystalline body \eq{10}, the $z_1$, $z_2$, $\ldots$, $z_n$ could be taken as the matrix elements (in some representation)
of the $m$ tensors $\BL_1$, $\BL_2$,$\ldots$, $\BL_m$. Alternatively we may just write $\BL(\Bx)=\BL(\Bx,\BL_1,\BL_2,\ldots,\BL_m)$ 
where it is implicitly understood that this is being regarded as a function of the elements of all the matrices $\BL_1$, $\BL_2$,$\ldots$,$\BL_m$ using
a basis of $\CT$  to represent these as matrices.

Let $\BGG_1^{\GO}$ denote the self-adjoint projection into $\CE_\GO^0$, which then annihilates any field in  $\overline\CJ_\GO$. Applying it to both sides of
\eq{6a} we get
\beq 0=\BM\BE-\BGG_1^{\GO}\Bh,\text{ where } \BM=\BGG_1^{\GO}\BL\BGG_1^{\GO}=\sum_{i=1}^nz_i\BM_i,\quad \BM^{(i)}=\BGG_1^{\GO}\BL^{(i)}\BGG_1^{\GO}. \eeq{7.a}
The generalized spectrum of $\BM$ can then be defined as the set of those values $\Bz=(z_1,z_2,\ldots,z_n)$ in $\mathbb{C}^n$ where the
operator $\BM$ does not have an inverse on the space $\CE_\GO^0$. Outside the generalized spectrum the inverse exists and defines the  Green's function $\BG_\GO$:
\beq \BE=\BG_\GO\Bh,\quad \BG_\GO=\BGG_1^{\GO}\BM^{-1}\BGG_1^{\GO}=\BGG_1^{\GO}(\BGG_1^{\GO}\BL\BGG_1^{\GO})^{-1}\BGG_1^{\GO}.
\eeq{7.b}
Note that there is an equivalence class of sources $\Bh$ that yield the same field $\BE$: we can add to $\Bh$ any field in $\overline\CJ_\GO$ without disturbing
$\BE$. In the context of the equations \eq{0} there is still only a unique $\BE(\Bx)$ associated with a given source $\Bf(\Bx)$, since if
$\Div\Bh'=\Div\Bh=\Bf$ then $\Bh'-\Bh\in\overline\CJ_\GO$.

We first show that $\BM$ has an inverse on $\CE_\GO^0$ if $\Bz=(z_1,z_2,\ldots,z_n)$ takes values in a domain $\CZ(\Ga,\Gb,\Gt)$ of $\mathbb{C}^n$, defined as 
that domain where $\BL$ satisfies the following boundedness and coercivity conditions:
\beq \Gb >\sup_{\substack{\BP\in\CH \\ |\BP|=1}}|\BL\BP|_\GO, \text{         (Boundedness)}
\eeq{8} 
\beq \Real (e^{i\Gt}\BL\BE,\BE)_\GO\geq \Ga|\BE|_\GO^2 \text{  for all } \BE\in\CE_{\GO}. \text{       (Coercivity)}
\eeq{9}
The latter is the usual coercivity condition and makes sense: if $\BE\in\CE_\GO^0$ is in the point spectrum of $\BM$ in the sense that $\BM\BE=0$ for some $z_1,z_2,\ldots,z_n$ then clearly
$(\BL\BE,\BE)_\GO=0$ and \eq{9} cannot hold. Under these boundedness and coercivity conditions, we will see that given any $\Bh$ there is a unique solution to \eq{6a} for $\BE$,
which thus defines the Green's operator $\BG_\GO$ in \eq{7.b}.

To show uniqueness, suppose for some given $\Bh$ that there is another solution $\BE'\in\CE_\GO^0$
and $\BJ'\in\overline\CJ_\GO$. Subtracting solutions we get
\beq \BJ-\BJ'=\BL(\BE-\BE'). \eeq{11a}
The coercivity condition \eq{9} with $\BP=\BE-\BE'$ implies
\beq \Ga|\BE-\BE'|^2=\Ga|\BP|^2\leq \Real (\BP,e^{i\Gt}\BL\BP) \leq \Real(\BE-\BE',e^{i\Gt}(\BJ-\BJ')).
\eeq{11b}
By the orthogonality of $\CE_\GO^0$ and $\overline\CJ_\GO$ the expression on the right is zero which forces $\BE=\BE'$,  thus establishing uniqueness.

To establish existence, we let $\BGG_1^{\GO}$ denote the projection into $\CE_\GO^0$,
which then annihilates any field in  $\overline\CJ_\GO$. Rewrite the constitutive law in \eq{4} as
\beq \BJ(\Bx)-c\BE(\Bx)=\Gd\BL(\Bx)\BE(\Bx)-\Bh(\Bx),\quad \Gd\BL(\Bx)=\BL(\Bx)-c\BI,
\eeq{14}
where $c=|c|e^{-i\Gt}$, and $|c|$ is a free constant that ultimately will be chosen very large to ensure convergence of a series expansion for $\BG_\GO$.
Applying the operator $\BGG_1^{\GO}/c$ to both sides, we get
 \beq -\BE=(\BGG_1^{\GO}\Gd\BL/c)\BE-\BGG_1^{\GO}\Bh/c
\eeq{15}
giving \eq{7.b} with
\beq \BG_\GO=[\BI+\BGG_1^{\GO}\Gd\BL/c]^{-1}\BGG_1^{\GO}/c
\eeq{18}
as our Green's function operator. Expanding this formulae, and using the fact that $(\BGG_1^{\GO})^2=\BGG_1^{\GO}$ we obtain the series expansion
\beq \BG_\GO=-(\BGG_1^{\GO}/c)\sum_{j=0}^\infty [\BGG_1^{\GO}(-\Gd\BL/c)\BGG_1^{\GO}]^j.
\eeq{19}
We want to show this converges to $\BG_\GO$ when $|c|$ is sufficiently large, and thus establishes the existence of the solution $\BE$ to \eq{6a} when
$\Bh$ is given.
Now, for any field $\BE\in\CE_\GO^0$, introduce $\BE'=-\BGG_1^{\GO}(\Gd\BL/c)\BE=\BGG_1^{\GO}(\BI-\BL/c)\BE$ giving $\BGG_1^{\GO}\BL\BE/c=\BE-\BE'$. The boundedness of $\BL$ and implies
\beq \Gb^2|\BE|^2/|c|^2\geq |\BE-\BE'|^2,
\eeq{20}
and by expanding $|\BE-\BE'|^2$ we get
\beq 2\Real (\BE,\BE')\geq |\BE'|^2+ [1-(\Gb/|c|)^2]|\BE|^2.
\eeq{21}
Now recall that $c$ has been chosen so $c/|c|=e^{-i\Gt}$. Then the coercivity \eq{9} implies
\beq
\Real (\BE,\BE-\BE') \geq \Ga|\BE|^2/|c|.
\eeq{22}
By combining this with \eq{21} we obtain
\beq [1+(\Gb/|c|)^2-2\Ga/|c|]|\BE|^2\geq |\BE'|^2.
\eeq{23}
Let
\beq  \|\BS\|_\GO= \sup_{\substack{\BP\in\CH \\ |\BP|_\GO=1}}|\BS\BP|_\GO \eeq{23.a}
denote the standard norm of an operator $\BS$. Then we get
\beqa &~&\|\BGG_1^{\GO}(\Gd\BL/c)\BGG_1^{\GO}\|_\GO^2 =   \sup_{\substack{\BE\in{\CE_\GO^0} \\ |\BE|_\GO=1}}(\BGG_1^{\GO}(\Gd\BL/c)\BE,\BGG_1^{\GO}(\Gd\BL/c)\BE)_\GO \nonum
&~& \quad\quad\quad \leq   1+(\Gb/|c|)^2-2\Ga/|c|=1-(\Ga/\Gb)^2 \text{    with }|c|=\Gb^2/\Ga. 
\eeqa{23.b}
Clearly this is less than $1$ and the series \eq{19} converges.

Incidentally, if we truncate the series expansion \eq{19}, the truncated expansion will be a polynomial
in the variables $z_1$, $z_2$, $\ldots$, $z_n$. Using the result that a sequence of analytic functions that converges uniformly on any compact set
of a domain is analytic in that domain [see theorem 10.28 of Rudin \cite{Rudin:1987:RCA}] we see
that $\BG_\GO$ is an operator valued analytic function of $\Bz$ in the domain $\CZ(\Ga,\Gb,\Gt)$ for any value of $\Gt$ and for arbitrarily small value of $\Ga>0$ and any arbitrarily large  value of
$\Gb$, such that \eq{8} and \eq{9} hold. This is also a corollary of the well known result that the Green's function is analytic
away from its spectrum. Clearly the domain of analyticity of $\BG_\GO(z_1,z_2,\ldots,z_n)$ is at least
\beq \bigcup_{\Gt,\Ga,\Gb}\CZ(\Ga,\Gb,\Gt).
\eeq{25}
Consequently the generalized spectrum must be confined to the set of $\mathbb{C}^n$ outside the region \eq{25}.

%%%%%%%%%%%%%%%%%%%%%%%%%%%%%%%%%%%%%%%%%%%%%%%%%%%%%%%%%%%%%%%%%%%%%%%%%%%%%%%%%%%%%%
\section{Herglotz type analytic properties of the Green's operator}
%\setcounter{equation}{0}
%%%%%%%%%%%%%%%%%%%%%%%%%%%%%%%%%%%%%%%%%%%%%%%%%%%%%%%%%%%%%%%%%%%%%%%%%%%%%%%%%%%%%%
In the theory of composites, the Herglotz type analytic property of the effective tensor
as a function of the moduli or tensors of the component materials has played an important role.
Particularly, it has served as a useful tool for deriving bounds on the effective tensor given the component
moduli, and possibly some information about the geometry such as the volume fractions
of the individual phases: see, for example, 
\cite{Bergman:1978:DCC,Milton:1979:TST,Milton:1981:BCP,Milton:1981:BTO,Bergman:1980:ESM,Bergman:1982:RBC,DellAntonio:1986:ATO,Golden:1983:BEP,Golden:1985:BEP,Clark:1994:MEC,Clark:1995:OBC,Clark:1997:CFR}
and Chapters 18, 27, and 28 of \cite{Milton:2002:TOC}. Recently, in Chapter 6 of \cite{Milton:2016:ETC} (available as https://arxiv.org/abs/1602.03383) Mattei and Milton found these analytic properties 
useful for deriving bounds on the response in the time domain. Also, it has been established by Cassier, Milton, and Welters that the Dirichlet to Neumann map governing the response of bodies shares
these Herglotz type properties: see Chapters 3 and 4 of \cite{Milton:2016:ETC}, the latter of which is also available as https://arxiv.org/abs/1512.05838.
Here we show that these Herglotz type properties extend to the Green's operator. Of course, analytic and Herglotz type properties of the Green's operator have an
extensive history: what is new here is the Herglotz type analytic properties as functions of the tensors $\BL_1$, $\BL_2$,$\dots$, $\BL_m$ of the phases.

Note that we are free to rewrite \eq{6a} as
\beq \BJ'(\Bx)=\BL'(\Bx)\BE(\Bx)-\Bh'(\Bx),\quad \BJ'\in\CJ,\quad \BE\in\CE,\quad \Bh'\in\CH,
\eeq{26}
with
\beq \BJ'=e^{i\Gy}\BJ,\quad \BL'=e^{i\Gy}\BL,\quad \Bh'=e^{i\Gy}\Bh.
\eeq{27}
So by appropriately redefining $\BL$ we may assume the coercivity condition \eq{9} is satisfied with $e^{i\Gt}=-i$, i.e. $\Gt=-\pi/2$. Then the condition implies
\beq
\Imag (\BP,\BL\BP)_\GO \geq (\BT\BP,\BP)_\GO+\Ga|\BP|_\GO^2\quad \text{   for all } \BP\in\CH_{\GO},
\eeq{28}
and since
\beq (\BG_\GO\Bh,\Bh)_\GO=(\BE,\Bh)_\GO=(\BE,\BL\BE-\BJ)_\GO=(\BE,\BL\BE)_\GO=(\BL\BE,\BL\BE)_\GO^*,
\eeq{29}
we conclude that for all $\Bh$ with support in $\GO$,
\beq \Imag (\Bh,\BG_\GO\Bh)_\GO=-\Imag(\BE,\BL\BE)_\GO\leq 0.
\eeq{30}
In the context of the $m$-phase polycrystalline material \eq{10} we may consider $\BG_\GO$ to be a function $\BG_\GO(\BL_1, \BL_2, \ldots,\BL_m)$
that is analytic in the domain where $\BL_1, \BL_2, \ldots,\BL_m$, and has the reverse Herglotz property
\beq \Imag (\Bh,\BG_\GO\Bh)_\GO\leq 0 \text{  when } \Imag\BL_i>0 \text{ for all }i,
\eeq{31}
the homogeneity property that for all $\Gl\in\mathbb{C}$,
\beq \BG_\GO(\Gl\BL_1, \Gl\BL_2, \ldots,\Gl\BL_m)=\Gl^{-1}\BG_\GO(\BL_1, \BL_2, \ldots,\BL_m),
\eeq{32}
and the normalization property that 
\beq \BG_\GO(\BI, \BI, \ldots,\BI)=\BGG_1^{\GO}.
\eeq{33}
Taking $\Gl=i$ in \eq{32}, and using \eq{30} we get
\beq \Real (\Bh,\BG_\GO\Bh)\geq 0 \text{  when } \Real\BL_i>0 \text{ for all }i.
\eeq{34}
\aj{Of course if the tensors $\BL_1$, $\BL_2$,\ldots,$\BL_m$ themselves have Herglotz properties as a function of frequency $\Go$, in the sense that 
their imaginary parts are positive definite when $\Go$ lies in the upper half plane $\Imag \Go>0$, then $\BG_\GO(\Go)$ will inherit the reverse
Herglotz property 
\beq  \Imag (\Bh,\BG_\GO\Bh)_\GO\leq 0 \text{  when } \Imag\Go>0, \eeq{34a}
and there will be associated integral representation formulas for the operator valued Herglotz function $-\BG_\GO(\Go)$, involving a positive semidefinite
operator valued measure derived from the values that $-\BG_\GO(\Go)$ takes as $\Go$ approaches the real axis.}

%%%%%%%%%%%%%%%%%%%%%%%%%%%%%%%%%%%%%%%%%%%%%%%%%%%%%%%%%%%%%%%%%%%%%%%%%%%%%%%%%%%%%%
\section{Using $Q_\GO$-convex functions to establish coercivity}
%\setcounter{equation}{0}
%%%%%%%%%%%%%%%%%%%%%%%%%%%%%%%%%%%%%%%%%%%%%%%%%%%%%%%%%%%%%%%%%%%%%%%%%%%%%%%%%%%%%%

The coercivity condition \eq{9} is generally hard to verify for a given operator $\BL$ as it requires one to test it for all fields $\BE\in\CE_\GO^0$.
However, suppose we are given a real valued quadratic form, $Q_\GO(\BP)$ defined for all $\BP\in\CH_\GO$, such that
\beq Q_\GO(\BE)\geq 0\quad\text{ for all }\BE\in\CE_\GO^0.
\eeq{Q.1}
If we can find a constant $t\geq 0$ such that the $Q_\GO$-coercivity condition defined by
\beq \Real (e^{i\Gt}\BL\BP,\BP)_\GO \geq tQ_\GO(\BE)+\Ga|\BP|_\GO^2 \quad\text{  for all } \BP\in\CH_{\GO},
\eeq{Q.2}
holds, then by taking $\BP=\BE$, with $\BE\in\CE_\GO^0$ it is clear that the coercivity condition \eq{9} will be satisfied.
Following ideas of Murat and Tartar \cite{Murat:1978:CPC,Tartar:1979:CCA,Tartar:1979:ECH,Murat:1981:CPC,Murat:1985:CVH,Tartar:1985:EFC} and Milton \cite{Milton:2013:SIG,Milton:2015:ATS} 
one can take this idea further. Given an integer $\ell\geq 1$, define 
\beqa 
\mathfrak{H}_\GO& = & \{\mathbb{P}~|~\mathbb{P}=(\BP_1,\BP_2,\ldots,\BP_\ell),~\BP_i\in\CH_\GO\text{ for }i=1,2,\ldots\ell\}, \nonum
\mathfrak{E}_\GO^0& = & \{\mathbb{E}~|~\mathbb{E}=(\BE_1,\BE_2,\ldots,\BE_\ell),~\BE_i\in\CE_\GO^0\text{ for }i=1,2,\ldots\ell\}.
\eeqa{Q.3}
Fields in these spaces take values in a tensor space $\mathfrak{T}$ consisting of tensors
\beq \mathfrak{T} = \{\mathbb{A}~|~\mathbb{A}=(\BA_1,\BA_2,\ldots,\BA_\ell),~\BA_i\in\CT \text{ for }i=1,2,\ldots\ell\},
\eeq{Q.3aa}
and on $\mathfrak{T}$ we define the inner product
\beq (\mathbb{A},\mathbb{A'})_{\mathfrak{T}}=\sum_{i=1}^\ell(\BA_i,\BA'_i)_{\CT}.
\eeq{Q.3ab}
Similarly, given any pair of fields $\mathbb{P},\mathbb{P}'\in \mathfrak{H}_\GO$ we define the inner product and norm:
\beq (\mathbb{P},\mathbb{P}')_{\mathfrak{H}_\GO}=\sum_{i=1}^\ell(\BP_i,\BP'_i)_{\GO},\quad |\mathbb{P}|_{\mathfrak{H}_\GO}=\sqrt{(\mathbb{P},\mathbb{P})_{\mathfrak{H}_\GO}}.
\eeq{Q.3a}
Introduce an operator $\mathbb{L}:\mathfrak{H}_\GO\to\mathfrak{H}_\GO$ defined by
\beq \mathbb{L}\mathbb{E}=\mathbb{L}(\BE_1,\BE_2,\ldots,\BE_\ell) =  (\BL\BE_1,\BL\BE_2,\ldots,\BL\BE_\ell).
\eeq{Q.4}
Now suppose we are given a real valued quadratic form, $Q_\GO(\mathbb{P})$ defined for all $\mathbb{P}\in\mathfrak{H}_\GO$, such that
\beq Q_\GO(\mathbb{E})\geq 0\text{ for all }\mathbb{E}\in\mathfrak{E}_\GO^0.
\eeq{Q.5}
If we can find a constant $t\geq 0$ such that the generalized $Q_\GO$-coercivity condition,
\beq \Real (e^{i\Gt}\mathbb{L}\mathbb{P},\mathbb{P})_{\mathfrak{H}_\GO} \geq t Q_\GO(\mathbb{P})+\Ga|\mathbb{P}|_{\mathfrak{H}_\GO}^2 
\text{  for all } \mathbb{P}\in\mathfrak{H}_{\GO}
\eeq{Q.6}
holds, then by taking $\mathbb{P}=\mathbb{E}$ where $\mathbb{E}\in\mathfrak{E}_\GO^0$ we get
\beq \Real (e^{i\Gt}\mathbb{L}\mathbb{E},\mathbb{E})_{\mathfrak{H}_\GO} \geq \Ga|\mathbb{E}|_{\mathfrak{H}_\GO}^2.
\eeq{Q.7}
This can be rewritten as
\beq \sum_{i=1}^\ell\Real(e^{i\Gt}\BL\BE_i,\BE_i)_{\GO}\geq \Ga\sum_{i=1}^\ell|\BE_i|^2_{\GO},
\eeq{Q.8}
and clearly implies that the coercivity condition \eq{9} holds. 
This choice of quadratic functions $Q_\GO$ enables one to get convergent series expansions for $\BG_\GO$ on a
domain $\CZ(\Ga,\Gb,\Gt,\ell,Q_\GO)$ defined as that region of $\mathbb{C}^n$ where the boundedness and $Q_\GO$-coercivity conditions 
\eq{8} and \eq{Q.6} hold, in which, by rescaling $Q_\GO$, one may assume that $t=1$. One should get tighter bounds on the generalized spectrum, as it lies outside the domain
\beq \bigcup_{\Gt,\Ga,\Gb,\ell,Q_\GO}\CZ(\Ga,\Gb,\Gt,\ell,Q_\GO).
\eeq{Q.11}
Obviously the $Q_\GO$-coercivity condition \eq{Q.2} is just a special case of the $Q_\GO$-coercivity condition \eq{Q.6} (obtained by taking $\ell=1$ or by
taking  $\mathbb{T}$ to be block diagonal). So from now on we will concentrate on finding quadratic functions $Q_\GO(\mathbb{P})$ satisfying \eq{Q.5}.
%%%%%%%%%%%%%%%%%%%%%%%%%%%%%%%%%%%%%%%%%%%%%%%%%%%%%%%%%%%%%%%%%%%%%%%%%%%%%%%%%%%%%%
\section{Finding appropriate quadratic $Q_\GO$-convex functions}
%\setcounter{equation}{0}
%%%%%%%%%%%%%%%%%%%%%%%%%%%%%%%%%%%%%%%%%%%%%%%%%%%%%%%%%%%%%%%%%%%%%%%%%%%%%%%%%%%%%%
\subsection{Generating position independent translations}
%%%%%%%%%%%%%%%%%%%%%%%%%%%%%%%%%%%%%%%%%%%%%%%%%%%%%%%%%%%%%%
Identifying quadratic forms such that $Q_\GO$-convexity condition \eq{Q.5} holds on $\mathfrak{E}_\GO^0$, and then determining 
when the $Q_\GO$-coercivity condition \eq{Q.6} holds is still a difficult task. Progress can be made by limiting attention
to quadratic functions $Q_\GO(\mathbb{P})$ generated by a self adjoint operator that acts locally in space, with action given by an operator $\mathbb{T}(\Bx)$, 
called the translation operator, so that
\beqa Q_\GO(\mathbb{P}) & = & \frac{1}{V(\GO)}\int_\GO(\mathbb{T}(\Bx)\mathbb{P}(\Bx),\mathbb{P}(\Bx))_{\mathfrak{T}}\,d\Bx \nonum
& = & \frac{1}{V(\GO)}\int_\GO\sum_{i=1}^\ell\sum_{j=1}^\ell(\BT_{ij}(\Bx)\BP_j(\Bx),\BP_i(\Bx))_\CT\,d\Bx,
\eeqa{F.0}
where the $\BT_{ij}(\Bx)$ represent the individual blocks of $\mathbb{T}(\Bx)$, with $\BT_{ji}(\Bx)=\BT_{ij}^\dagger(\Bx)$ to ensure that $\mathbb{T}$ is self-adjoint.
The key point is that $\mathbb{T}$ need not act separately on the individual components of $\mathbb{P}$, but can couple them. Then the $Q_\GO$-coercivity
condition \eq{Q.6} becomes
\beq  \int_\GO\sum_{i=1}^\ell\Real(e^{i\Gt}\BL(\Bx)\BP_i(\Bx),\BP_i(\Bx))_{\CT}\,d\Bx\geq t\int_\GO\sum_{i=1}^\ell\sum_{j=1}^\ell(\BT_{ij}(\Bx)\BP_j(\Bx),\BP_i(\Bx))_{\CT}\,d\Bx
+\Ga\int_\GO\sum_{i=1}^\ell|\BP_i|^2_{\CT}\,d\Bx.
\eeq{F.0a}
and will clearly be satisfied if and only if for all $\Bx\in\GO$, and for all $\mathbb{A}=(\BA_1,\BA_2,\ldots,\BA_\ell)\in\mathfrak{T}$,
\beq \sum_{i=1}^\ell\Real(e^{i\Gt}\BL(\Bx)\BA_i,\BA_i)_{\CT}\geq t\sum_{i=1}^\ell\sum_{j=1}^\ell(\BT_{ij}(\Bx)\BA_j,\BA_i)_{\CT}
+\Ga\sum_{i=1}^\ell|\BA_i|^2_{\CT}.
\eeq{F.0b}
This can be written as a block matrix inequality:
\beq \bpm \Real(e^{i\Gt}\BL(\Bx))-\Ga\BI & 0 & \ldots & 0 \\
                0 & \Real(e^{i\Gt}\BL(\Bx))-\Ga\BI & \ldots &0 \\
                \vdots & \vdots & \ddots & \vdots \\
                0 & 0 & \ldots & \Real(e^{i\Gt}\BL(\Bx))-\Ga\BI\epm
                -t\bpm \BT_{11}(\Bx) & \BT_{12}(\Bx) & \ldots & \BT_{1\ell}(\Bx) \\
                \BT_{21}(\Bx) & \BT_{22}(\Bx) & \ldots & \BT_{2\ell}(\Bx) \\
                \vdots & \vdots & \ddots & \vdots \\
                \BT_{\ell 1}(\Bx) & \BT_{\ell 2}(\Bx) & \ldots & \BT_{\ell\ell}(\Bx)\epm\geq 0,
\eeq{F.0c}
where the inequality holds in the sense that the matrix on the left is a positive definite matrix. Thus, when the moduli are shifted (translated) by $\mathbb{T}(\Bx)$,
they satisfy a local coercivity condition on $\mathfrak{T}$.

 For a multicomponent medium with $\BL(\Bx)$ given
by \eq{10A} this inequality clearly becomes
\beq \bpm \Real(e^{i\Gt}\BL_i)-\Ga\BI & 0 & \ldots & 0 \\
                0 & \Real(e^{i\Gt}\BL_i)-\Ga\BI & \ldots &0 \\
                \vdots & \vdots & \ddots & \vdots \\
                0 & 0 & \ldots & \Real(e^{i\Gt}\BL_i)-\Ga\BI\epm
                -t\bpm \BT_{11}(\Bx) & \BT_{12}(\Bx) & \ldots & \BT_{1\ell}(\Bx) \\
                \BT_{21}(\Bx) & \BT_{22}(\Bx) & \ldots & \BT_{2\ell}(\Bx) \\
                \vdots & \vdots & \ddots & \vdots \\
                \BT_{\ell 1}(\Bx) & \BT_{\ell 2}(\Bx) & \ldots & \BT_{\ell\ell}(\Bx)\epm\geq 0,
\eeq{F.0d}
for all $\Bx$ in phase $i$ (i.e. such that $\chi_i(\Bx)=1$), and for all
$i=1,2,\ldots,n$.

In the case where $\mathbb{T}$ is a constant operator, independent of $\Bx$, there is a simple algebraic route to finding $\mathbb{T}$ for which
the $Q_\GO$-convexity condition \eq{Q.5} is satisfied. Then the inequalities in \eq{Q.5} will surely hold if they hold on the larger space
$\mathfrak{E}$, where $\mathfrak{E}$ is comprised of fields $\mathbb{E}(\Bx)=(\BE_1(\Bx),\BE_2(\Bx),\ldots,\BE_\ell(\Bx))$
defined for $\Bx\in\mathbb{R}^d$ where all the component fields $\BE_i$ lie in $\CE$. Thus we need to identify quadratic functions $Q_*(\mathbb{A})$ defined for $\mathbb{A}\in\mathfrak{T}$ 
that satisfy
\beq \int_{\mathbb{R}^d} Q_*(\mathbb{E}(\Bx))\,d\Bx\geq 0 \text{ for all }\mathbb{E}\in\mathfrak{E}.
\eeq{F.1}
We call this the $Q_*$-convexity condition, and functions that satisfy it will be called $Q_*$-convex. The terminology arises because of similarities with
the definition of $Q_*$-convexity given in \cite{Milton:2013:SIG,Milton:2015:ATS}, where it was used to obtain boundary field inequalities. 
Then we may simply take
\beq Q_\GO(\mathbb{E})=\int_\GO~ Q_*(\mathbb{E}(\Bx))\,d\Bx.
\eeq{F.2}
The field $\mathbb{E}(\Bx)$, being square integrable, will have some Fourier expansion
\beq \mathbb{E}(\Bx)=\int_\mathbb{R}^d\widehat{\mathbb{E}}(\Bk)e^{i\Bk\cdot\Bx}\,d\Bk,
\eeq{F-3}
and the differential constraints on $\mathbb{E}(\Bx)$ imply that $\widehat{\mathbb{E}}(\Bk)$ takes values in
some subspace $\mathfrak{E}_\Bk$. Substituting \eq{F-3} in \eq{F.1}, and doing the integration over $\Bx$ we see that
\eq{F.1} holds if
\beq  Q_*(\mathbb{E}(\Bk))=(\mathbb{E}(\Bk)\mathbb{T},\mathbb{E}(\Bk))_{\mathfrak{T}}\geq 0\text{ for all }\Bk\in\mathbb{R}^d.
\eeq{F-4}
Thus the $Q_*$-convexity condition \eq{F.1} is satisfied if the algebraic condition that
\beq Q_*(\mathbb{A})\geq 0 \text{ for all } \mathbb{A}\in\mathfrak{E}_\Bk
\eeq{F.3}
holds for all $\Bk\in\mathbb{R}^d$. As $\mathbb{T}$ is selfadjoint, if we make the substitution $\mathbb{A}= \mathbb{A}'+i\mathbb{A}''$,
where $\mathbb{A}'$ and $\mathbb{A}''$ are the real and imaginary parts of $\mathbb{A}$, we obtain
\beq Q_*(\mathbb{A})=Q_*(\mathbb{A}')+Q_*(\mathbb{A}'').
\eeq{F.3a}
Now suppose the spaces $\CH$, $\CE$ and $\CJ$ are real-symmetric, in the sense that
if a field belongs to them, so does the complex conjugate field. Then if $\mathbb{A}$ lies in $\mathfrak{E}_\Bk$,
so too does its complex conjugate, and hence $\mathbb{A}'\in\mathfrak{E}_\Bk$ and $\mathbb{A}''\in\mathfrak{E}_\Bk$.
With this assumption on the subspaces we see from \eq{F.3a} that it is only necessary to test \eq{F.3}
for real values of $\mathbb{A}\in\mathfrak{E}_\Bk$. 

The condition \eq{F.3} for $Q_*$-convexity reduces to
the normal quasiconvexity condition when the subspace $\CE$ is scale invariant, i.e. if $\BE\in\CE$, then
the field $\BE'(\Bx)$ defined by $\BE'(\Bx)=\BE(\Gl\Bx)$ also lies in $\CE$ for all real nonzero choices of the scale factor
$\Gl$.  In that event, $\mathfrak{E}_\Bk$ remains invariant under the scale change $\Bk\to\Bk/\Gl$. 

When $\mathbb{T}$ does not depend on $\Bx$ then the inequalities \eq{F.0d} do not involve the characteristic
functions of the phases and thus we obtain geometry independent bounds on the generalized spectrum (in the sense that these 
bounds hold for all choices of the $\chi_i(\Bx)$). For polycrystalline multiphase bodies where $\BL(\Bx)$
takes the form \eq{10}, one has to made additional assumptions about $\mathbb{T}$ to obtain geometry independent
bounds that are independent both of the choice of the $\chi_i(\Bx)$ and the rotation field $\BR(\Bx)$. Here we
follow the ideas developed in \cite{Avellaneda:1988:ECP}: see also sections 24.2 and 24.8 in \cite{Milton:2002:TOC}.
It is convenient to
take $\ell$ as the dimension of $\CT$. Then, taking an orthonormal basis $\Be_1$, $\Be_2$, $\ldots$, $\Be_\ell$ of $\CT$ we can regard
the elements of $\mathfrak{T}$ as linear maps from $\CT$ to $\CT$. Thus if $\mathbb{A}=(\BA_1,\BA_2,\ldots,\BA_\ell)\in\mathfrak{T}$ is
given, we define its action on $\CT$ through its action on the basis elements according to the prescription
\beq \mathbb{A}\Be_i=\BA_i. \eeq{FF.1}
Now given a rotation $\BR$ we can define an associated operator $\mathbb{R}(\BR)$ acting on  linear maps from $\CT$ to $\CT$, defined
by 
\beq [\mathbb{R}(\BR)\mathbb{A}]\Be_j=\CR(\BR)^\dagger\mathbb{A}[\CR(\BR)\Be_j], \eeq{FF.2}
where $\CR(\BR):\CT\to\CT$ is that operator on the tensor space which corresponds to a rotation $\BR$: see \eq{5b}. 
Since $\CR(\BR)\Be_j$ is a linear combination of the $\Be_i$, the action of $\mathbb{A}$ on $\CR(\BR)\Be_j$ can be computed via
\eq{FF.1}. Now $\mathbb{T}$ acts on a map $\mathbb{A}$ from  $\CT$ to $\CT$ to produce a new map $\mathbb{T}\mathbb{A}$ from $\CT$ to $\CT$.
We seek $\mathbb{T}$ that are rotationally invariant in the sense that for all  rotations $\BR$,
\beq \mathbb{T}[\mathbb{R}(\BR)\mathbb{A}]=\mathbb{R}(\BR)[\mathbb{T}\mathbb{A}].
\eeq{FF.3}
Now $\mathbb{L}$ defined in \eq{Q.4}, acts locally and we can write
\beq [\mathbb{L}\mathbb{P}](\Bx)=[\mathbb{L}(\Bx)]\mathbb{P}(\Bx)=\BL(\Bx)\mathbb{P}(\Bx),
\eeq{FF.4}
which thus defines $\mathbb{L}(\Bx)$ where the expression on the far right can be regarded as the composition of the two
operators $\BL(\Bx):\CT\to\CT$ and $\mathbb{P}(\Bx):\CT\to\CT$. Similarly, we can define the operator $\widetilde{\mathbb{L}}$ that 
acts locally, given by
\beq [\widetilde{\mathbb{L}}\mathbb{P}](\Bx)=[\widetilde{\mathbb{L}}(\Bx)]\mathbb{P}(\Bx)=\widetilde{\BL}(\Bx)\mathbb{P}(\Bx)
\eeq{FF.4a}
in which $\widetilde{\BL}(\Bx)$ is defined by \eq{10}. Then we have
\beqa [\mathbb{L}\mathbb{R}(\BR)\mathbb{A}]\Be_j & = & \BL\CR(\BR)^\dagger\mathbb{A}[\CR(\BR)\Be_i] \nonum
& = & \CR(\BR)^\dagger\widetilde{\BL}\mathbb{A}[\CR(\BR)\Be_i] \nonum
& = & \mathbb{R}(\BR)[\widetilde{\mathbb{L}}]\Be_j.
\eeqa{FF.4b}

The condition \eq{Q.6} then holds if and only if 
\beq \Real (e^{i\Gt}\mathbb{L}\mathbb{A},\mathbb{A})_{\mathfrak{T}} \geq t (\mathbb{T}\mathbb{A},\mathbb{A})_{\mathfrak{T}}+\Ga(\mathbb{A},\mathbb{A})_{\mathfrak{T}} 
\text{  for all } \mathbb{A}:\CT\to\CT.
\eeq{FF.5}
We are free to replace $\mathbb{A}$ by $\mathbb{R}(\BR)\mathbb{A}$ and then using the rotational invariance of $\mathbb{T}$, and the rotational
properties \eq{FF.4b} of $\mathbb{L}$, the condition becomes
\beq \Real (e^{i\Gt}\widetilde{\mathbb{L}}\mathbb{A},\mathbb{A})_{\mathfrak{T}} \geq t (\mathbb{T}\mathbb{A},\mathbb{A})_{\mathfrak{T}}+\Ga(\mathbb{A},\mathbb{A})_{\mathfrak{T}} 
\text{  for all } \mathbb{A}:\CT\to\CT.
\eeq{FF.6}
With $\widetilde{\BL}(\Bx)$ given by \eq{10} we see \eq{FF.6} holds if and only if \eq{F.0d} is satisfied.
%%%%%%%%%%%%%%%%%%%%%%%%%%%%%%%%%%%%%%%%%%%%%%%%%%%%%%%%%%%%%%%%%%%%%%%%%%%%%%%%%%%%%%%%%%%%%%%%%%%%
\subsection{Generating position dependent translations using Null-Lagrangians or coordinate transformations}
%%%%%%%%%%%%%%%%%%%%%%%%%%%%%%%%%%%%%%%%%%%%%%%%%%%%%%%%%%%%%%%%%%%%%%%%%%%%%%%%%%%%%%%%%%%%%%%%%%%%%
More general $Q_\GO$ functions, with $\mathbb{T}(\Bx)$ depending on $\Bx$, can also be obtained. The simplest case is when $\mathbb{E}(\Bx)$ 
is a linear function of a potential $\Bu(\Bx)$, its gradient $\Grad\Bu(\Bx)$. Then we may look for quadratic functions $L(\Bx,\Bu,\Grad\Bu)$ that are Null-Lagrangians
in the sense that
\beq \int_{\GO}L(\Bx,\Bu(\Bx),\Grad\Bu(\Bx))\,d\Bx =0.
\eeq{F.3aa}
Such functions have been completely characterized, even if they are not quadratic \cite{Olver:1988:SNL}. 
The integrand here may not be expressible simply as a quadratic function of $\mathbb{E}(\Bx)$ but we could look for operators $\mathbb{T}(\Bx)$
such that for all  $\Bx\in\GO$,
\beq (\mathbb{T}(\Bx)\mathbb{E}(\Bx),\mathbb{E}(\Bx))_{\mathfrak{T}}\geq L(\Bx,\Bu(\Bx),\Grad\Bu(\Bx)),
\eeq{F.3b}
where here $\Bu(\Bx)$ and $\Grad\Bu(\Bx)$ are those arguments that appear implicitly in the expression for $\mathbb{E}(\Bx)$. 

Another approach is to make coordinate transformations.
Thus, for example, suppose we replace $\Bx$ by $\By$, and suppose the differential constraints on 
$\mathbb{E}(\By)$ imply that is a linear function of a potential $\Bw(\By)$, its gradient $\Grad_{\By}\Bw(\By)$, a field $\BV(\By)$ and
its divergence $\nabla_{\By}\cdot\BV(\By)$ (that is possibly constrained to be zero). Here $\Grad_{\By}$ and $\nabla_{\By}\cdot$ denote the gradient 
and divergence with respect to the $\By$ variables. Furthermore, suppose we have identified 
a quadratic function $Q_*(\mathbb{E}(\By))=F(\Bw(\By),\Grad_{\By}\Bw(\By),\BV(\By),\nabla_{\By}\cdot\BV(\By))$ such that 
\beq \int_{\GO}F(\Bw(\By),\Grad_{\By}\Bw(\By),\BV(\By),\nabla_{\By}\cdot\BV(\By))\,d\By=\int_{\GO} Q_*(\mathbb{E}(\By))\,d\By\geq 0 \text{ for all }\mathbb{E}\in\mathfrak{E}.
\eeq{F.4}
Now consider a coordinate transformation from $\By$ to a variable $\Bx$ such that $\By=\BGy(\Bx)$, i.e., $\Bx=\BGy^{-1}(\By)$
that leaves points outside and on the boundary of $\GO$ unmoved, i.e.
\beq \psi(\By)=\By\text{  for  }\By\in\Md\GO \text{ or }\By\text{  outside } \GO.
\eeq{F.5}
Then making the change of variables in the integral \eq{F.5} we obtain
\beq \int_{\GO}\widetilde{F}(\Bx,\Bu(\Bx),\Grad\Bu(\Bx),\BQ(\Bx),\Div\BQ(\Bx))\,d\By \geq 0,
\eeq{F.6}
where with $\By=\BGy(\Bx)$,
\beqa &~& \Bu(\Bx)=\Bw(\By), \quad \BQ(\Bx)=\BGL(\By)\BV(\By)/\det[\BGL(\By)], \nonum
 & ~& \Grad\Bu(\Bx)=[\BGL(\By)]^{-1}\Grad_{\By}\Bw(\By),\quad \Div\BQ(\Bx)=[\nabla_{\By}\cdot\BV(\By)]/\det[\BGL(\By)],\nonum                               \nonum
&~& \GL_{ij}(\By)=\frac{\Md x_j(\By)}{\Md y_i}=\frac{\Md \Gy^{-1}_j(\By)}{\Md y_i},
\eeqa{F.6a}
and
\beqa
&~& \widetilde{F}(\Bx,\Bu(\Bx),\Grad\Bu(\Bx),\BQ(\Bx),\Div\BQ(\Bx)) \nonum
&~&\quad\quad =\frac{F(\Bu(\Bx),\BGL(\BGy(\Bx))\Grad\Bu(\Bx),\det[\BGL(\BGy(\Bx))][\BGL(\BGy(\Bx))]^{-1}\BQ(\Bx),\det[\BGL(\BGy(\Bx))]\Div\BQ(\Bx))}{\det[\BGL(\BGy(\Bx))]}, \nonum
&~&
\eeqa{F.7}
in which $\Div\BQ=0$ if $\nabla_{\By}\cdot\BV(\By)=0$. We now can take
\beq Q_\GO(\widetilde{\mathbb{E}})=\int_{\GO}\widetilde{F}(\Bx,\Bu(\Bx),\Grad\Bu(\Bx),\BQ(\Bx),\Div\BQ(\Bx))\,d\Bx, \eeq{F.7a}
where we suppose that $\widetilde{\mathbb{E}}$ is a linear function of the potential $\Bu(\By)$, its gradient $\Grad_{\By}\Bu(\By)$, and a divergence free field $\BQ(\By)$,
and involves them in such a way that the integrand in \eq{F.7a} can be expressed simply as a function of $\widetilde{\mathbb{E}}(\Bx)$ (otherwise, one would
look for a quadratic function of $\widetilde{\mathbb{E}}(\Bx)$ that bounds $\widetilde{F}(\Bx,\Bu(\Bx),\Grad\Bu(\Bx),\BQ(\Bx),\Div\BQ(\Bx))$ from above for all $\Bx\in\GO$,
in a similar fashion as was done in \eq{F.3b}).
%%%%%%%%%%%%%%%%%%%%%%%%%%%%%%%%%%%%%%%%%%%%%%%%%%%%%%%%%%%%%%%%%%%%%%%%%%%%%%%%%%%%%%%%%%%%%%%%%%%%
\subsection{Generating position dependent translations by making substitutions}
%%%%%%%%%%%%%%%%%%%%%%%%%%%%%%%%%%%%%%%%%%%%%%%%%%%%%%%%%%%%%%%%%%%%%%%%%%%%%%%%%%%%%%%%%%%%%%%%%%%%%

Yet another approach is to make substitutions. As suggested by the analysis of \cite{Olver:1988:SNL}
having identified a quadratic function  $Q_*(\mathbb{E}(\Bx))=F(\Bw(\Bx),\Grad\Bw(\Bx))$ such that
\beq \int_{\GO}F(\Bw(\Bx),\Grad\Bw(\Bx))\,d\Bx=\int_{\GO} Q_*(\mathbb{E}(\Bx))\,d\Bx\geq 0 \text{ for all }\mathbb{E}\in\mathfrak{E},
\eeq{F.8}
we make the substitution
\beq \Bw(\Bx)=\BS(\Bx)\Bu(\Bx)+\Bs(\Bx),
\eeq{F.9}
where we require that $\Bs(\Bx)=0$ and $\Bu(\Bx)=0$ for $\Bx\in\Md\GO$ to ensure that $\Bw(\Bx)=0$ on $\Md\GO$. Here we suppose that at each $\Bx\in\GO$, $\Bw(\Bx)$
and $\Bs(\Bx)$ are $p$-component vectors, while $\Bu(\Bx)$ is an $s$-component vector, so $\BS(\Bx)$ is a $p\times s$ matrix. 
Since $F$ is quadratic we may write 
\beq F(\Bw,\BW)=w_\Ga A_{\Ga\Gb}w_\Gb^*+2\Real\left(w _\Ga B_{\Ga j\Gb}W_{j\Gb}^*\right)+W_{i\Ga}C_{i\Ga j\Gb}W_{j\Gb}^* ,
\eeq{F.10}
where sums over repeated indices are assumed (Einstein summation convention) and
$\BA$ and $\BC$ are Hermitian ($A_{\Ga\Gb}^*=A_{\Gb\Ga}$ and $C_{i\Ga j\Gb}^*=C_{j\Gb i\Ga}$).
Making the substitution \eq{F.9} we get
\beqa &~& F(\BS(\Bx)\Bu(\Bx)+\Bs(\Bx),[\Grad\Bu(\Bx)]\BS(\Bx)^T+[\Grad\BS(\Bx)]\Bu(\Bx)+\Grad\Bs(\Bx)) \nonum
&~&\quad\quad= F(\BS(\Bx)\Bu(\Bx),[\Grad\Bu(\Bx)]\BS(\Bx)^T+[\Grad\BS(\Bx)]\Bu(\Bx))+F(\Bs(\Bx),\Grad\Bs(\Bx))\nonum
&~&\quad\quad\quad +2\Real[u_\Gg(S_{\Gb\Gg}A_{\Gb\Ga}s_\Ga^*+S_{\Ga\Gg}B_{\Ga j \Gb}s_{\Gb,j}^*+s_\Ga^*B_{\Ga j \Gb}^*S_{\Gb\Gg,j}+S_{\Ga\Gg,i}C_{i\Ga j\Gb}s_{\Gb,j}^*)] \nonum
&~&\quad\quad\quad +2\Real[u_{\Gg,i}(S_{\Gb\Gg}B_{\Ga i \Gb}^*s_{\Ga}^*+S_{\Ga\Gg} C_{i\Ga j\Gb} s_{\Gb,j}^*)].
\eeqa{F.11}
Let us require that for all $\Gg=1,2,\ldots,s$,
\beqa &~& S_{\Gb\Gg}A_{\Gb\Ga}s_\Ga^*+S_{\Ga\Gg}B_{\Ga j \Gb}s_{\Gb,j}^*+s_\Ga^*B_{\Ga j \Gb}^*S_{\Gb\Gg,j}+S_{\Ga\Gg,i}C_{i\Ga j\Gb}s_{\Gb,j}^*=0,\nonum
&~& (S_{\Gb\Gg}B_{\Ga i \Gb}^*s_{\Ga}^*+S_{\Ga\Gg} C_{i\Ga j\Gb} s_{\Gb,j}^*)_{,i}=0.
\eeqa{F.13}
Given $\Bs(\Bx)$ this places $2s$ linear restrictions on the $ps$ components of $\BS(\Bx)$. So one would expect to find solutions, at least when $p\geq 2$.
Then the function
\beqa \widetilde{F}(\Bx,\Bu(\Bx),\Grad\Bu(\Bx))& = &  
F(\BS(\Bx)\Bu(\Bx),[\Grad\Bu(\Bx)]\BS(\Bx)^T+[\Grad\BS(\Bx)]\Bu(\Bx)) \nonum
& = &F(\BS(\Bx)\Bu(\Bx)+\Bs(\Bx),[\Grad\Bu(\Bx)]\BS(\Bx)^T+[\Grad\BS(\Bx)]\Bu(\Bx)+\Grad\Bs(\Bx)) \nonum
&~&-F(\Bs(\Bx),\Grad\Bs(\Bx))-2\Real[u_{\Gg,i}(S_{\Gb\Gg}B_{\Ga i \Gb}^*s_{\Ga}^*+S_{\Ga\Gg} C_{i\Ga j\Gb} s_{\Gb,j}^*)]
\eeqa{F.14}
is a quadratic function of $\Bu$ and $\Grad\Bu$. In general the identity \eq{F.8} holds as an inequality. However there can be special
fields $\Bw(\Bx)$ such that it holds as an equality \cite{Kang:2013:BVF3d,Milton:2013:SIG,Milton:2015:ATS}. Special fields
are more likely to exist if $Q_*(\mathbb{E}(\Bx))$ is chosen to be extremal in the sense that it loses its $Q_*$-convexity property
if any strictly convex quadratic function is subtracted from it. In the case where $\mathbb{T}(\Bx)$ does not depend on $\Bx$,
an algorithm for numerically constructing extremal $Q_*$-convex functions (based on earlier work in \cite{Kohn:1988:OBE,Milton:1990:CSP}) 
is given in \cite{Milton:2013:SIG}.
For functions of gradients, an explicit example of an extremal $Q_*$-convex function that is not a Null-Lagrangian is given
in \cite{Harutyunyan:2015:EEE} and curiously there is a connection between such functions and extremal polynomials \cite{Harutyunyan:2015:REE,Harutyunyan:2016:TCE}.

Let $\Bs(\Bx)$ be one of these special fields. Then we have
\beq \int_\GO F(\Bs(\Bx),\Grad\Bs(\Bx))\,d\Bx=0,
\eeq{F.15}
and, using the fact that $\Bu$ vanishes on $\Md\GO$, we see that
\beqa &~& \int_\GO \widetilde{F}(\Bx,\Bu(\Bx),\Grad\Bu(\Bx))\,d\Bx \nonum
&~&\quad\quad =\int_\GO F(\Bw(\Bx),\Grad\Bw(\Bx))\,d\Bx
-\int_\GO2\Real[u_{\Gg,i}(S_{\Gb\Gg}B_{\Ga i \Gb}^*s_{\Ga}^*+S_{\Ga\Gg} C_{i\Ga j\Gb} s_{\Gb,j}^*)] \nonum
&~&\quad\quad \geq -\Real\int_{\Md\GO}u_\Gg(S_{\Gb\Gg}B_{\Ga i \Gb}^*s_{\Ga}^*+S_{\Ga\Gg} C_{i\Ga j\Gb} s_{\Gb,j}^*)n_i \nonum
&~&\quad\quad \geq 0,
\eeqa{F.16}
in which the $n_i(\Bx)$ are the elements of $\Bn(\Bx)$, the outwards normal to $\Md\GO$.

Substitutions can also be made if $\mathbb{E}(\Bx)$ involves a field $\BV(\Bx)$ and its divergence $\Bv(\Bx)=\Div\BV(\Bx)$. Having identified a quadratic function 
$Q_*(\mathbb{E}(\By))=F(\BV(\By),\Div\BV(\By))$ such that
\beq \int_{\GO}F(\BV(\Bx),\Div\BV(\Bx))\,d\Bx=\int_{\GO} Q_*(\mathbb{E}(\Bx))\,d\Bx\geq 0 \text{ for all }\mathbb{E}\in\mathfrak{E},
\eeq{F.17}
we make the substitution
\beq \BV(\Bx)=\BQ(\Bx)\BS(\Bx)+\Bs(\Bx).
\eeq{F.18}
Since $F$ is quadratic we may write 
\beq F(\BV,\Bv)=V_{i\Ga} A_{i\Ga j\Gb}V_{j\Gb}^*+2\Real\left(v_\Ga B_{\Ga j\Gb}V_{j\Gb}^*\right)+v_{\Ga}C_{\Ga\Gb}v_{\Gb}^* ,
\eeq{F.19}
where $\BA$ and $\BC$ are Hermitian ($A_{i\Ga j\Gb}^*=A_{j\Gb i\Ga}$ and $C_{\Ga\Gb}^*=C_{\Gb\Ga}$).
Making the substitution \eq{F.9} we get
\beqa &~& F(\BS(\Bx)\BQ(\Bx)+\Bs(\Bx),[\Div\BQ(\Bx)]\BS(\Bx)+[\BQ(\Bx)]^T\Grad\BS(\Bx)+\Div\Bs(\Bx)) \nonum
&~&\quad\quad= F(\BS(\Bx)\BQ(\Bx),[\Div\BQ(\Bx)]\BS(\Bx)+[\BQ(\Bx)]^T\Grad\BS(\Bx)) +F(\Bs(\Bx),\Div\Bs(\Bx))\nonum
&~&\quad\quad\quad +2\Real[Q_{i\Gg}(S_{\Gg\Ga}A_{i\Ga j\Gb}s_{j\Gb}^*+S_{\Gg\Ga,i}B_{\Ga j \Gb}^*s_{j\Gb}^*+S_{\Gg\Gb}B_{\Ga i \Gb}^*s_{j\Ga,j}^*+S_{\Gg\Ga,i}C_{\Ga\Gb}s_{j\Gb,j}^*)] \nonum
&~&\quad\quad\quad +2\Real[Q_{i\Gg,i}(S_{\Gg\Ga}B_{\Ga j \Gb}^*s_{j\Gb}^*+S_{\Gg\Ga} C_{\Ga\Gb} s_{j\Gb,j}^*)].
\eeqa{F.20}
We require that
\beqa &~& S_{\Gg\Ga}A_{i\Ga j\Gb}s_{j\Gb}^*+S_{\Gg\Ga,i}B_{\Ga j \Gb}^*s_{j\Gb}^*+S_{\Gg\Gb}B_{\Ga i \Gb}^*s_{j\Ga,j}^*+S_{\Gg\Ga,i}C_{\Ga\Gb}s_{j\Gb,j}^*)=Y_{\Gg,i}, \nonum
&~& S_{\Gg\Ga}B_{\Ga j \Gb}^*s_{j\Gb}^*+S_{\Gg\Ga} C_{\Ga\Gb} s_{j\Gb,j}^*=0,
\eeqa{F.21}
for some potential $\BY$. Then the function
\beqa  \widetilde{F}(\BQ(\Bx),\Div\BQ(\Bx)) & = & F(\BS(\Bx)\BQ(\Bx),[\Div\BQ(\Bx)]\BS(\Bx)+[\BQ(\Bx)]^T\Grad\BS(\Bx)) \nonum
& = & F(\BS(\Bx)\BQ(\Bx)+\Bs(\Bx),[\Div\BQ(\Bx)]\BS(\Bx)+[\BQ(\Bx)]^T\Grad\BS(\Bx)+\Div\Bs(\Bx)) \nonum
&~&\quad-F(\Bs(\Bx),\Div\Bs(\Bx)) -2\Real[Q_{i\Gg}Y_{\Gg,i}]
\eeqa{F.22}
is a quadratic function of $\BQ(\Bx)$ and $\Div\BQ(\Bx)$. Again we look for special fields $\Bs(\Bx)$ such that 
\beq \int_\GO F(\Bs(\Bx),\Div\Bs(\Bx))\,d\Bx=0.
\eeq{F.23}
We get
\beqa \int_\GO\widetilde{F}(\BQ(\Bx),\Div\BQ(\Bx))\,d\Bx & = &\int_\GO F(\BV(\Bx),\Div\BV(\Bx))\,d\Bx-2\Real[Q_{i\Gg}Y_{\Gg,i}] \nonum
&\geq & \int_{\Md\GO}Y_{\Gg}Q_{i\Gg}n_i\,dS=0,
\eeqa{F.24}
where we have used the fact that $\Bn\cdot\BQ=0$ on $\GO$, as implied by the boundary constraint $\Md\mathbb{E}=0$.

 %%%%%%%%%%%%%%%%%%%%%%%%%%%%%%%%%%%%%%%%%%%%%%%%%%%%%%%%%%%%%%%%%%%%%%%%%%%%%%%%%%%%%%
\section{Periodic boundary conditions}
%\setcounter{equation}{0}
%%%%%%%%%%%%%%%%%%%%%%%%%%%%%%%%%%%%%%%%%%%%%%%%%%%%%%%%%%%%%%%%%%%%%%%%%%%%%%%%%%%%%%

The theory we have developed can easily be extended to allow for other boundary conditions. The extension
to Green's functions where we impose the constraint that $\Md\BJ=0$ on the boundary of $\GO$, rather than
$\Md\BE=0$, is obvious: just swap the roles of the spaces $\CE$ and $\CJ$. 

Periodic boundary conditions, with $\GO$ as the unit cell of periodicity, have the advantage that one has an explicit expression in Fourier space for the projection operator $\BGG_1^\GO$
onto $\CE_\GO$ which is taken as the space of all $\GO$-periodic fields, that are square integrable in the unit cell of periodicity and satisfy the appropriate differential constraints.
$\CJ_\GO$ is then taken as the orthogonal complement of $\CE_\GO$ in the space $\CH_\GO$ of $\GO$-periodic fields that are square integrable within the unit cell of periodicity.
Thus we also have explicit expression in Fourier space for the projection operator $\BGG_2^\GO=\BI-\BGG_1^\GO$ onto the space $\CJ_\GO$. We are then interested in solutions
of \eq{6a} as before, with $\BL(\Bx)$ being $\GO$-periodic. The analysis proceeds as before, and with $\mathfrak{E}_\GO^0$ defined by \eq{Q.3},
we now can directly identify functions $Q_\GO(\mathbb{E})$ satisfying \eq{Q.5} when they are of the form \eq{F.0}
with $\mathbb{T}(\Bx)$ independent of $\Bx$.

Further bounds on the generalized spectrum of the Green's operator can be obtained when $\BL(\Bx)$ is non-singular for all $\Bx\in\GO$. Then \eq{6a} can be rewritten as 
\beq  \BE(\Bx)=[\BL(\Bx)]^{-1}\BJ(\Bx)-\widetilde{\Bh},\quad \BE\in\CE_\GO,\quad\BJ\in\CJ_\GO,\quad \widetilde{\Bh}=-[\BL(\Bx)]^{-1}\Bh(\Bx)\in\CH_\GO.
\eeq{P.1}
This is exactly the same form as before but with the roles of $\CE_\GO$ and $\CJ_\GO$, and $\Bh$ and $\widetilde{\Bh}$ interchanged, and with $\BL(\Bx)$ replaced by
$[\BL(\Bx)]^{-1}$. If $\widetilde{\BG}_\GO$ is the Green's operator for this problem, so that $\BJ=\widetilde{\BG}_\GO\widetilde{\Bh}$, we clearly have $\BE=\BG_\GO\Bh$ with
\beq \BG_\GO=[\BL(\Bx)]^{-1}-[\BL(\Bx)]^{-1}\widetilde{\BG}_\GO[\BL(\Bx)]^{-1}.
\eeq{P.2}
So if, for some $\widetilde{\Gb}>\widetilde{\Ga}>0$, 
the boundedness and coercivity conditions
\beqa &~&  \widetilde{\Gb} >\sup_{\substack{\BP\in\CH \\ |\BP|=1}}|\BL^{-1}\BP|_\GO, \nonum 
 &~& \Real (e^{i\Gt}\BL^{-1}\BJ,\BJ)_\GO\geq \widetilde{\Ga}|\BJ|_\GO^2 \text{  for all } \BJ\in\CJ_{\GO},
\eeqa{P.3}
are met (and $\BL(\Bx)$ is non-singular for all $\Bx\in\GO$), then $\widetilde{\BG}_\GO$ and hence $\BG_\GO$ exists.

More generally, following ideas developed in \cite{Cherkaev:1994:VPC}, section 18 of \cite{Milton:1990:CSP}, \cite{Milton:2009:MVP,Milton:2010:MVP}
and Chapter 14 of \cite{Milton:2016:ETC}, we may take a splitting of the operator $\BL$:
\beq \BL=\BL_A+\BL_B, \eeq{P.4}
in which $\BL_A$ is non-singular. One may, for example, take $\BL_A$ as the Hermitian part of $\BL$ and  $\BL_B$
as the anti-Hermitian part, but other choices are possible too. 
Then we consider the pair of equations
\beqa \BJ & = & [\BL_A+\BL_B]\BE-\Bh,\quad \BJ\in\CJ_\GO,\quad \BE\in\CE_\GO,\quad\Bh\in\CH_\GO,\nonum
 \BJ' & = & [\BL_A-\BL_B]\BE'-\Bh',\quad \BJ'\in\CJ_\GO,\quad \BE'\in\CE_\GO,\quad\Bh'\in\CH_\GO,
\eeqa{P.5}
the first of which is of course equivalent to \eq{6a}. Adding and subtracting these gives
\beqa \BJ+\BJ'& = & \BL_A(\BE+\BE')+\BL_B(\BE-\BE')-\Bh-\Bh', \nonum
\BJ-\BJ'& = & \BL_A(\BE-\BE')+\BL_B(\BE+\BE')-\Bh+\Bh', \eeqa{P.6}
or equivalently, in block matrix form,
\beq \bpm  \BJ+\BJ' \\ \BJ-\BJ'\epm
=\bpm \BL_A & ~&  \BL_B \\  \BL_B  &~&  \BL_A \epm
\bpm \BE+\BE' \\ \BE-\BE'\epm-\bpm \Bh+\Bh' \\ \Bh-\Bh'\epm.
\eeq{P.7}
Going one step further, we can solve the first equation in \eq{P.6} for $\BE+\BE'$ in terms of
$\BJ+\BJ'$ and $\BE-\BE'$:
\beq \BE+\BE'=\BL_A^{-1}[\BJ+\BJ'-\BL_B(\BE-\BE')+\Bh+\Bh']. \eeq{P.8}
Then substituting this in the second equation in \eq{P.6} gives
\beq \BJ-\BJ'=\BL_A(\BE-\BE')+\BL_B\BL_A^{-1}[\BJ+\BJ'-\BL_B(\BE-\BE')+\Bh+\Bh']-\Bh+\Bh'.
\eeq{P.9}
These two equations take the block matrix form
\beqa \underbrace{\bpm c_E(\BE+\BE') \\ c_J(\BJ-\BJ') \epm}_{\underline{\BJ}}
& = &  \underbrace{\bpm c_Ed_J\BL_A^{-1} & ~ & -c_Ed_E\BL_A^{-1}\BL_B \\ c_Jd_J\BL_B\BL_A^{-1} & ~& c_Jd_E(\BL_A-\BL_B\BL_A^{-1}\BL_B) \epm}_{\underline{\BL}}
\underbrace{\bpm (\BJ+\BJ')/d_J \\ (\BE-\BE')/d_E \epm}_{\underline{\BE}} \nonum
&~&\quad\quad -\underbrace{\bpm -c_E\BL_A^{-1}(\Bh+\Bh') \\ c_J(\Bh-\Bh') -c_J\BL_B\BL_A^{-1}(\Bh+\Bh') \epm}_{\underline{\Bh}}. 
\eeqa{P.10}
in which we have introduced the additional nonzero complex factors $c_E$, $c_J$, $d_E$, and $d_J$ as they may help in establishing coercivity.
We now define 
\beqa \underline{\CH}_\GO & = &\left\{\underline{\BP}=\left.\bpm \BP_1 \\ \BP_2 \epm\quad \right| \quad\BP_1,\BP_2\in \CH_\GO\right\}, \nonum
\underline{\CE}_\GO & = & \left\{\underline{\BE}=\left.\bpm \BJ_0 \\ \BE_0 \epm \quad \right| \quad\BJ_0 \in \CJ_\GO,\quad\BE_0\in \CE_\GO\right\}, \nonum
\underline{\CJ}_\GO & = &\left\{\underline{\BE}=\left.\bpm \BE_0 \\ \BJ_0 \epm \quad \right| \quad\BE_0 \in \CE_\GO,\quad\BJ_0\in \CJ_\GO\right\},
\eeqa{P.11}
and on $\underline{\CH}_\GO$ we introduce the inner product and norm
\beq \left(\bpm \BP_1 \\ \BP_2 \epm,\bpm \BP_3 \\ \BP_4 \epm\right)_{\underline{\CH}_\GO}=(\BP_1,\BP_3)_\GO+(\BP_2,\BP_4)_\GO,
\quad |\underline{\BP}|_{\underline{\CH}_\GO}=\sqrt{(\underline{\BP},\underline{\BP})_{\underline{\CH}_\GO}},
\eeq{P.11a}
so that the subspaces $\underline{\CE}_\GO$ and $\underline{\CJ}_\GO$ are orthogonal complements. 
If, for some $\underline{\Gb}> \underline{\Ga}>0$,  $\underline{\BL}$ satisfies the boundedness and coercivity conditions, 
\beqa &~&  \underline{\Gb} >\sup_{\substack{\underline{\BP}\in\underline{\CH}_\GO \\ |\underline{\BP}|=1}}|\underline{\BL}\,\underline{\BP}|_{\underline{\CH}_\GO}, \nonum 
 &~& \Real (\underline{\BL}\,\underline{\BE},\underline{\BE})_{\underline{\CH}_\GO}\geq\underline{\Ga}|\underline{\BE}|_{\underline{\CH}_\GO}^2 \text{  for all } \underline{\BE}\in\underline{\CE}_{\GO},
\eeqa{P.12}
then, say with $\Bh'=-\Bh$, \eq{P.10}  will have a unique solution
for the fields $\underline{\BE}$ and $\underline{\BJ}$. From these fields one extracts the fields $\BE$, $\BJ$, $\BE'$ and $\BJ'$ that solve \eq{P.5}. Conversely,
if the equations \eq{P.5} did not have a unique solution, there would not be a unique solution to \eq{P.10} in contradiction to what the coercivity of $\underline{\BL}$ implies.
Thus if the boundedness and coercivity conditions \eq{P.12} hold, then the Green's operator $\BG_\GO$ for the equations \eq{6a} exists. As previously, one can look for appropriate
$Q_\GO$-convex functions that guarantee that the coercivity condition in \eq{P.12} holds. At this stage it is unclear
if sharper bounds on the generalized spectrum can be obtained by making this splitting -- this needs further exploration.  

 %%%%%%%%%%%%%%%%%%%%%%%%%%%%%%%%%%%%%%%%%%%%%%%%%%%%%%%%%%%%%%%%%%%%%%%%%%%%%%%%%%%%%%
\section{Quasiperiodic Periodic boundary conditions and bounds on the Floquet-Bloch spectrum}
%\setcounter{equation}{0}
%%%%%%%%%%%%%%%%%%%%%%%%%%%%%%%%%%%%%%%%%%%%%%%%%%%%%%%%%%%%%%%%%%%%%%%%%%%%%%%%%%%%%%

Quasiperiodic boundary conditions, as appropriate to wave equations, can easily be dealt with too within the existing framework,
and the associated spectrum is the Floquet-Bloch spectrum. Bounds on this spectrum for periodic operators are of wide interest. Following the seminal papers of 
John \cite{John:1987:SLP} and Yablonovitch \cite{Yablonovitch:1987:ISE} there was tremendous excitement in obtaining 
acoustic, electromagnetic and elastic band-gap materials for which the relevant operator had no spectrum in a frequency window
for any value of $\Bk$. A rigorous proofs of the existence of band-gaps was given by Figotin and Kuchment \cite{Figotin:1996:BGSI,Figotin:1996:BGSII}.
See also the book of Joannopoulos et.al. \cite{Joannopoulos:2008:ISS}. Upper bounds on photonic band gaps have been given by 
Rechtsman and Torquato \cite{Rechtsman:2009:MOU}. More recently Lipton and Viator \cite{Lipton:2017:BWC} obtained convergent power series
for the Bloch wave spectrum in periodic media. For general theory related to Floquet theory 
for partial differential equations, and Bloch waves, see Wilcox's paper \cite{Wilcox:1978:TBW} and Kuchment's book \cite{Kuchment:1993:FTP}.
 
Here we concentrate on the acoustic equations \eq{5.A} as
the extension to other wave equations is obvious. We assume that the effective mass density matrix $\BGr(\Bx,\Go)$, and the bulk modulus
$\Gk(\Bx,\Go)$ are periodic with now $\GO$ denoting the unit cell of periodicity. We look for solutions where
the pressure $\BP(\Bx)$ and velocity $\Bv(\Bx)$ are quasiperiodic, i.e. 
\beq \BP(\Bx)=e^{i\Bk\cdot\Bx}\widetilde{\BP}(\Bx),\quad \Bv(\Bx)=e^{i\Bk\cdot\Bx}\widetilde{\Bv}(\Bx),
\eeq{B-1}
in which $\widetilde{\BP}(\Bx)$ and $\widetilde{\Bv}(\Bx)$ are periodic functions of $\Bx$ with period cell $\GO$. It follows that
\beqa \Grad\BP(\Bx)& = & e^{i\Bk\cdot\Bx}\Grad\widetilde{\BP}(\Bx)+i\Bk e^{i\Bk\cdot\Bx}\Bk\widetilde{\BP}(\Bx), \nonum
\Div\Bv(\Bx)& = & e^{i\Bk\cdot\Bx}\Div\widetilde{\Bv}(\Bx)+ie^{i\Bk\cdot\Bx}\Bk\cdot\widetilde{\Bv}(\Bx).
\eeqa{B-2}
Making this substitution in \eq{5.A} we obtain
\beq \begin{pmatrix}-i\widetilde{\Bv} \\ \Bk\cdot\widetilde{\Bv}-i\Div\widetilde{\Bv} \end{pmatrix}
=\begin{pmatrix}-(\Go\BGr)^{-1} & 0 \\ 0 & \Go/\Gk\end{pmatrix}\begin{pmatrix}i\Bk\widetilde{P}+\Grad\widetilde{P} \\  \widetilde{P}\end{pmatrix},
\eeq{B-3}
or equivalently
\beq \underbrace{\begin{pmatrix}-i\widetilde{\Bv} \\-i\Div\widetilde{\Bv} \end{pmatrix}}_{\widetilde{\BJ}(\Bx)}
=\underbrace{\begin{pmatrix}-(\Go\BGr)^{-1} & -i(\Go\BGr)^{-1}\Bk \\ i\Bk(\Go\BGr)^{-1} & z+(\Go/\Gk)\end{pmatrix}}_{\widetilde{\BL}(\Bx)}\underbrace{\begin{pmatrix} \Grad\widetilde{P} \\  \widetilde{P}\end{pmatrix}}_{\widetilde{\BE}(\Bx)},
\eeq{B-4}
where $z=\Bk\cdot\Bk$. Written in this form we see that $\widetilde{\BL}$ is a linear function of $z$ and the elements of $\Bk$, and hence we recover
the known result \cite{Wilcox:1978:TBW} that the Green's function will be analytic in these variables away from the spectrum. 
All the analysis applies as before. We take $\CE_{\GO}$ to consist of $\GO$-periodic fields $\widetilde{\BE}(\Bx)$ having the same form
as that appearing on the left hand side of \eq{B-4} and we take $\CJ_{\GO}$ to consist of $\GO$-periodic fields $\widetilde{\BJ}(\Bx)$ having the same form
as that appearing on the right hand side of \eq{B-4}. Integration shows that these spaces are orthogonal. 
%%%%%%%%%%%%%%%%%%%%%%%%%%%%%%%%%%%%%%%%%%%%%%%%%%%%%%%%%%%%%%%%%%%%%%%%%%%%%%%%%%%%%%%%%%%%%%%%%%%%%%%%%%%%%%%%%%%%%%%%5
\acknowledgements{\aj{The author is grateful to Yury Grabovsky for pointing out that in the formulation of \eq{6a} one needs the closure
$\overline\CJ_\GO$ of $\CJ_\GO$ rather than just $\CJ_\GO$. Also the referee is thanked for helpful comments.}  The author is thankful to the National Science Foundation for support through the Research Grant DMS-1211359.}
%\bibliography{/home/milton/tcbook,/home/milton/newref}
%

%%%%%%%%%%%%%%%%%%%%%%%%%%%%%%%%%%%%%%%%%%%%%%%%%%%%%%%%%%%%%%%%%%%
\section*{Appendix: Canonical forms for some other equations in the frequency domain }
%%%%%%%%%%%%%%%%%%%%%%%%%%%%%%%%%%%%%%%%%%%%%%%%%%%%%%%%%%%%%%%%%%%%%%%%%%%%%
In Chapter 1 of \cite{Milton:2016:ETC} canonical forms of various linear equation  of physics were suggested. For thermoelasticity and vibrating plates
these were formulated in the time domain. Here we give the associated canonical forms of these equations in the frequency domain. In addition
we review, and correct, the canonical form for the equations of thermoacoustics in the frequency domain. A canonical form for these equations 
was provided in section 1.8 of \cite{Milton:2016:ETC}, but the analysis contained some errors (notably, the fields were assumed to have a $e^{-i\Go t}$ time dependence
while the source material for the equations assumed a $e^{i\Go t}$ time dependence). Here we take the opportunity to correct those errors. The source
used for the equations of linearized thermoacoustics, which
incorporate thermal and viscous losses, was a COMSOL
Acoustics Module User's Guide (\cite{COMSOL:2013:AMU}), equations (7-5), page 286. Some related theory can be found in
\cite{Pierce:1981:AIP}.

The time-harmonic linearized thermoacoustic equations involve the density fluctuations, temperature fluctuations, pressure fluctuations, stress, and velocity fields
which are the real parts of $e^{-i\Go t}\Gr(\Bx)$, $e^{-i\Go t}\Gt(\Bx)$, $e^{-i\Go t} P(\Bx)$, $e^{-i\Go t} \BGs(\Bx)$, and $e^{-i\Go t} \Bv(\Bx)$,
where the complex fields, in the absence of source terms, satisfy
\beq i\Go\Gr =  -\Gr_0(\Div\Bv), \quad
 i\Go\Gr_0\Bv  =  \Div \BGs, \quad i\Go(\Gr_0C_p\Gt-T_0\Ga_0P)=\Div[k(\Bx)\Grad \Gt],
\eeq{*1}
representing the equations of conservation of mass, momentum, and energy: where $\Gr_0$ and $T_0$ are the background density
and temperature; $C_p$ is the heat capacity
at constant pressure; $\Ga_0$ is the coefficient of thermal expansion
at constant pressure; and $k(\Bx)$ is the thermal conductivity. 
Additionally one has the relations
\beqa \Gr & = & \Gr_0(\Gb_TP-\Ga_0\Gt), \quad  \BGs=-P\BI+\BCD\Grad\Bv, \nonum
\BCD\Grad\Bv & = & \Gm[\Grad\Bv+(\Grad\Bv)^T]+\left(\Gm_B-\frac{2}{3}\Gm\right)(\Div\Bv)\BI,
\eeqa{A0.26b}
where the first equation is the linearization of the equation of state
linking pressure, density, and temperature, while the
second and third equations give the constitutive law for the stress in a fluid, in terms the velocity gradient and pressure fields. Here
$\BCD$ is the isotropic fourth-order tensor of viscosity moduli,
$\Gm$ and $\Gm_B$ are the dynamic shear and bulk viscosities
(see \cite{Dukhin:2009:BVC} for a discussion of bulk viscosity)
and $\Gb_T$ is the isothermal compressibility.
We first eliminate the density $\Gr$ from these equations to get an expression for the pressure $P$ in terms of the other variables:
\beq P=\frac{-i}{\Go\Gb_T}\Div\Bv +\frac{\Ga_0 \Gt}{\Gb_T}, \eeq{A0.26d}
which we can use to eliminate $P$ from the other equations:
\beqa  \BGs & = & \BCD\Grad\Bv+\frac{i\BI}{\Go\Gb_T}\Div\Bv -\frac{\Ga_0 \Gt\BI}{\Gb_T}, \nonum
\Div[k(\Bx)\Grad \Gt] &= &
-i\Go\Gr_0C_p\Gt+\frac{\Ga_0T_0}{\Gb_T}\Div\Bv +i\Go\frac{\Ga_0^2T_0 \Gt}{\Gb_T}.
\eeqa{A0.26e}
Introducing the heat flux $\Bq=-k(\Bx)\Grad \Gt$, we can rewrite the thermoacoustic equations (without sources) as
\beq
\underbrace{\begin{pmatrix}
i\BGs \\
i\Div\BGs \\
i\Bq \\
i\Div\Bq 
\end{pmatrix}}_\BJ
=
\underbrace{\begin{pmatrix}
i\BCD(\Bx)-\frac{\BI\otimes\BI}{\Go\Gb_T} & 0 & 0 & \frac{i\Ga_0T_0\BI}{\Gb_T} \\
0 & \Go\Gr_0 & 0 & 0  \\
0 & 0 & ik(\Bx)T_0 & 0  \\
\frac{-i\Ga_0T_0\BI\cdot}{\Gb_T} & 0 & 0 & \Go\Gr_0C_pT_0\Gb_0/\Gb_T
\end{pmatrix}}_{\BL(\Bx)}
\underbrace{\begin{pmatrix}
\Grad\Bv \\
\Bv \\
-\Grad \Gt/T_0 \\
-\Gt/T_0 
\end{pmatrix}}_{\BE(\Bx)}.
\eeq{*2}
where
\beq \Gb_0=\Gb_T-\frac{\Ga_0^2T_0}{\Gr_0C_p} \eeq{*2.1}
is the adiabatic compressibility. As desired,
the matrix $\BL(\Bx)$ entering this constitutive law is such that the Hermitian part of $\BL/i$ is positive semidefinite when $\Imag\Go\geq 0$. Furthermore, we have the key identity,
\beq
(\BJ(\Bx),\BE(\Bx))_\CT=\begin{pmatrix}
i\BGs \\
i\Div\BGs \\
i\Bq \\
i\Div\Bq
\end{pmatrix}\cdot
\begin{pmatrix}
\Grad\Bv^* \\
\Bv^* \\
-\Grad \Gt^*/T_0 \\
-\Gt^*/T_0
\end{pmatrix}
=i\Div[\BGs\Bv^*-\Bq\Gt^*/T_0],
\eeq{*3}
and  $\Md\BE$ can then be identified with the boundary values of $\Bv$ and $\Gt/T_0$ while 
$\Md\BJ$ can be identified with the boundary values of $\BGs\cdot\Bn$ and $\Bq\cdot\Bn$, in which
$\Bn$ is the normal to $\Md\GO$.

The canonical equations of thermoelasticity in the frequency domain take a similar form.
These equations (\cite{Chandrasekharaiah:1986:ACT}; see also \cite{Norris:1994:DGF}) take the form
\beqa -\Gr\Go^2 & = & \Div\BGs, \quad \Gs_{ij}=C_{ijk\ell}\Ge_{k\ell}-\Gb_{ij}\Gt, \nonum
\Gr S& = & (\Gr c/\Gt_0)\Gt+\Gb_{ij}\Ge_{ij}, \quad \Div\Bq-i\Go\Gt_0\Gr S=0, \quad
\Bq=-\BGk(\Go)\Grad\Gt
\eeqa{A0.68a}
where $\BGs$ is the stress; $\BGe=[\Grad\Bu+(\Grad\Bu)^T]$ is the strain;
$\Bq$ is the heat flux; $S$ is the entropy change; $\Gt$ is the change in temperature above the ambient
temperature $\Gt_0$; $c$ is the specific heat per unit mass at constant temperature; the $\Gb_{ij}=\Gb_{ji}$ are essentially
coefficients of thermal expansion, $\Bk(\Go)$ is the (matrix-valued) thermal conductivity tensor. In general, $\BGk(\Go)$
is frequency dependent and complex if there is some thermal relaxation. By eliminating $S$, the thermoelasticity equations can now be written as
\beq
\underbrace{\begin{pmatrix}
i\BGs\\
i\Div\BGs \\
i\Bq \\
i\Div\Bq
\end{pmatrix}}_{\BJ(\Bx)}
=\underbrace{\begin{pmatrix}
-\BCC/\Go & 0 & 0 & i\BGb T_0 \\
0 & \Go\Gr & 0 & 0  \\
0 & 0 & i T_0 \BGk(\Go) & 0  \\
-i\BGb T_0 & 0 & 0  & \Go T_0\Gr c \\
\end{pmatrix}}_{\BL(\Bx)}
\underbrace{\begin{pmatrix}
-i\Go\Grad\Bu\\
-i\Go\Bu \\
-\Grad\Gt/ T_0 \\
-\Gt/ T_0
\end{pmatrix}}_{\BE(\Bx)}.
\eeq{*4}
Again, as desired,
the matrix $\BL(\Bx)$ entering this constitutive law is such that the Hermitian part of $\BL/i$ is positive semidefinite when $\Imag\Go\geq 0$.
The key identity,
\beq (\BJ(\Bx),\BE(\Bx))_\CT=\begin{pmatrix}
i\BGs\\
i\Div\BGs\\
i\Bq\\
i\Div\Bq
\end{pmatrix}\cdot\begin{pmatrix}
i\Go^*\Grad\Bu^*\\
i\Go^*\Bu^* \\
-\Grad\Gt^*/ T_0 \\
-\Gt^*/ T_0
\end{pmatrix}
=-\Div(\Go^*\BGs\Bu^*+i\Bq\Gt^*/ T_0),
\eeq{*5}
holds and $\Md\BE$ can then be identified with the boundary values of $\Go\Bu$ and $\Gt/T_0$ while 
$\Md\BJ$ can be identified with the boundary values of $\BGs\cdot\Bn$ and $\Bq\cdot\Bn$, in which
$\Bn$ is the normal to $\Md\GO$.

For thin plates, the dynamic plate equations at constant frequency
can be written in the form
\beq
\underbrace{\begin{pmatrix}
i\BM\\
\nabla\cdot(\nabla\cdot\BM)
\end{pmatrix}}_{\BJ(\Bx)}
=
\underbrace{\begin{pmatrix}
-\BCD(\Bx)/\Go & 0 \\
0 & h(\Bx)\Go\Gr(\Bx)
\end{pmatrix}}_{\BL(\Bx)}
\underbrace{\begin{pmatrix}
\nabla\nabla v
\\
i v
\end{pmatrix}}_{\BE(\Bx)},
\eeq{*6}
where $\BM(\Bx,t)$ is the bending moment tensor,
$\BCD(\Bx)$ is the fourth-order tensor
of plate rigidity coefficients, $h(\Bx)$ is the plate thickness, $\Gr(\Bx)$ is the density,
and $v=\Md w/\Md t$ is the velocity of the vertical deflection $w(\Bx,t)$ of the plate.
Note that the matrix $\BL(\Bx)$ has positive definite imaginary part when $\Go$ has positive imaginary part, and we have the key identity,
\beq (\BJ(\Bx),\BE(\Bx))_\CT=\begin{pmatrix}
i\BM\\
\nabla\cdot(\nabla\cdot\BM)
\end{pmatrix}\cdot\begin{pmatrix}
\nabla\nabla v
\\
-i v
\end{pmatrix}=i\Div[\BM\cdot\Grad v-v\Div\BM].
\eeq{*7}
So in the context of these equations,  $\Md\BE$ can be identified with the boundary values of $\Grad v$ and $v$ while
$\Md\BJ$ can be identified with the boundary values of $\BM\Bn$ and $(\Div\BM)\cdot\Bn$, in which $\Bn$
is the outwards normal to $\Md\GO$. 

For moderately thick plates we need to replace these equations by those of \cite{Mindlin:1951:IRI} (see also \cite{Larsen:2009:TML}).
We now assume the plate material is locally isotropic. With a $e^{-\Go t}$ time dependence, the equilibrium equations
for a small plate element read as
\beqa -\Gr h^3\Go^2\Gy_x/12 & = & T_x-M_{x,x}-M_{xy,y}, \nonum
-\Gr h^3\Go^2\Gy_y/12 & = & T_y-M_{y,y}-M_{xy,x}, \nonum
-\Gr h\Go^2 w & = & T_{x,x}+T_{y,y}.
\eeqa{A0.37fa}
Here $w$ is the out of plane deflection; $\Gy_x$ and $\Gy_y$ are the angles of rotation; $M_x$, $M_y$ and $M_{xy}$ are the bending moments (we use the abbreviated
notation $M_x$ and $M_y$ to denote the components $M_{xx}$ and $M_{yy}$ of the tensor field $\BM(\Bx)$); $T_x$ and $T_y$ are the shear forces;
$\Gr=\Gr(x,y)$ is the plate density; and $h=h(x,y)$ is the plate thickness. 
The relation which links the bending moments and shear forces to the deflection and
rotation angles takes the form
\beq  \begin{pmatrix} M_x \\ M_y \\ M_{xy} \\ T_x \\T_y \end{pmatrix}=-\BD\begin{pmatrix} \Gy_{x,x} \\ \Gy_{y,y} \\ \Gy_{x,y}+\Gy_{y,x} \\ \Gy_x-w_{,x} \\ \Gy_y-w_{,y} \end{pmatrix},
\eeq{*8}
where the subscript comma denotes differentiation and the stiffness matrix $\BD$ is
\beq \BD=\BD(x,y)=\begin{pmatrix} D & \Gv D & 0 & 0 & 0 \\
                         \Gv D & D & 0 & 0 & 0 \\
                         0 & 0 & \frac{1-\Gv}{2}D & 0 & 0 \\
                         0 & 0 & 0 & kGh & 0 \\
                         0 &  0 & 0 & 0 & kGh \end{pmatrix}.
\eeq{*9}
Here $D=Eh^3/[12(1-\Gv^2)]$ is the flexural rigidity; $E$ is the Young modulus; $G$ is the shear modulus; $\Gv$ is the Poisson ratio; and $k$ is a shear correction factor taking the value $5/6$ for a
plate. We now can rewrite the equations in the canonical form:
\beq \underbrace{\begin{pmatrix} -i{M}_{x} \\ -i{M}_{y} \\ -i{M}_{xy} \\ -i{T}_{x} \\ -i{T}_{y}\\ T_x-M_{x,x}-M_{xy,y} \\ T_y-M_{y,y}-M_{xy,x} \\ T_{x,x}+T_{y,y} \end{pmatrix}}_\BJ
=\BL(x,y)\underbrace{\begin{pmatrix} -i\Go{\Gy}_{x,x} \\ -i\Go{\Gy}_{y,y} \\ -i\Go{\Gy}_{x,y}-i\Go{\Gy}_{y,x} \\ -i\Go{\Gy}_x+i\Go{w}_{,x} \\ -i\Go{\Gy}_y+i\Go{w}_{,y})
\\ -\Go{\Gy}_{x} \\ -\Go{\Gy}_{y} \\ -\Go{w}
 \end{pmatrix}}_\BE,
\eeq{*10}
where the matrix $\BL(x,y)$ entering the
constitutive law takes the block-diagonal Hermitian form
\beq \BL(x,y)=\begin{pmatrix} -\BD(x,y)/\Go & 0 & 0 & 0 \\
                               0 & \Go\Gr(x,y)[h(x,y)]^3/12 & 0 & 0 \\
                               0 & 0 & \Go\Gr(x,y)[h(x,y)]^3/12 & 0 \\
                               0 & 0 & 0 & \Go\Gr(x,y)h(x,y) \end{pmatrix}.
\eeq{*11}
Again, the matrix $\BL(\Bx)$ has positive definite imaginary part when $\Go$ has positive imaginary part, and we have the key identity
\beqa (\BJ(\Bx),\BE(\Bx))_\CT &= & \begin{pmatrix} -i{M}_{x} \\ -i{M}_{y} \\ -i{M}_{xy} \\ -i{T}_{x} \\ -i{T}_{y}\\ T_x-M_{x,x}-M_{xy,y} \\ T_y-M_{y,y}-M_{xy,x} \\ T_{x,x}+T_{y,y} \end{pmatrix}
\cdot\begin{pmatrix} i\Go^*{\Gy}_{x,x}^* \\ i\Go^*{\Gy}_{y,y}^* \\ i\Go^*{\Gy}_{x,y}^*+i\Go^*{\Gy}_{y,x}^* \\ i\Go^*{\Gy}_x^*-i\Go^*{w}_{,x}^* \\ i\Go^*{\Gy}_y^*-i\Go^*{w}_{,y}^*
\\ \Go^*{\Gy}_{x}^* \\ \Go^*{\Gy}_{y}^* \\ \Go^*{w}^*
 \end{pmatrix}\nonum
& = & \frac{\Md}{\Md x}[{M}_{x}\Go^*{\Gy}_{x}^*-{T}_{x}\Go^*{w}^*+{M}_{xy}\Go^*{\Gy}_{y}^*]
+\frac{\Md}{\Md y}[{M}_{y}\Go^*{\Gy}_{y}^*-{T}_{y}\Go^*{w}^*+{M}_{xy}\Go^*{\Gy}_{x}^*].\nonum &~&
\eeqa{*12}
Consequently, $\Md\BE$ can be identified with the boundary values of $\Go{\Gy}_{x}$, $\Go{\Gy}_{y}$ and $\Go w$ while
$\Md\BJ$ can be identified with the boundary values of $\BM\Bn$ and $n_x{T}_{x}+n_y{T}_{y}$, in which $\Bn=(n_x,n_y)$
is the outwards normal to $\Md\GO$.

%$\Bu(\Bx,t)$ is the displacement in the solid phase, $\Bw(\Bx,t)$ is the relative fluid displacement, $\BGs$ is the
%stress in the solid, $P$ is the fluid pressure,
%$\Gr$ and $\Gr_f$ are the solid and fluid densities, the $\widehat{m}_{ij}(\Go)$ are viscodynamic terms
%Hence the Biot equations can be rewritten as
%\beq
%\underbrace{\begin{pmatrix}
%-i\BGs\\
%\Div\BGs \\
%-\Grad P \\
%i\Go P \\
%-i\Go M\Gz
%\end{pmatrix}}_{\BJ(\Bx)}
%=\underbrace{\begin{pmatrix}
%-\BCC/\Go & 0 & 0 & \BM & 0 \\
%0 & \Go\Gr & \Go\Gr_f & 0 & 0 \\
%0 & \Go\Gr_f & \Go\widehat{m}_{ij}(\Go) & 0 & 0 \\
%\BM & 0 & 0 & \Go M & \Go M \\
%0 & 0 & 0 & \Go M & \Go M
%\end{pmatrix}}_{\BL(\Bx)}
%\underbrace{\begin{pmatrix}
%-\Grad\Bv\\
%-i\Bv \\
%-i\Bw \\
%i\Div\Bw \\
%s/\Go
%\end{pmatrix}}_{\BE(\Bx)}.
%\eeq{*14}
%\beq (\BJ(\Bx),\BE(\Bx))_\CT=
%\eeq{*15}

%\bibliographystyle{/home/milton/reports/nsf/bibstyle/x-unsrt}
%\bibliographystyle{plain}
%\bibliography{/home/milton/tcbook,/home/milton/newref}

\begin{thebibliography}{86}%
\makeatletter
\providecommand \@ifxundefined [1]{%
 \@ifx{#1\undefined}
}%
\providecommand \@ifnum [1]{%
 \ifnum #1\expandafter \@firstoftwo
 \else \expandafter \@secondoftwo
 \fi
}%
\providecommand \@ifx [1]{%
 \ifx #1\expandafter \@firstoftwo
 \else \expandafter \@secondoftwo
 \fi
}%
\providecommand \natexlab [1]{#1}%
\providecommand \enquote  [1]{``#1''}%
\providecommand \bibnamefont  [1]{#1}%
\providecommand \bibfnamefont [1]{#1}%
\providecommand \citenamefont [1]{#1}%
\providecommand \href@noop [0]{\@secondoftwo}%
\providecommand \href [0]{\begingroup \@sanitize@url \@href}%
\providecommand \@href[1]{\@@startlink{#1}\@@href}%
\providecommand \@@href[1]{\endgroup#1\@@endlink}%
\providecommand \@sanitize@url [0]{\catcode `\\12\catcode `\$12\catcode
  `\&12\catcode `\#12\catcode `\^12\catcode `\_12\catcode `\%12\relax}%
\providecommand \@@startlink[1]{}%
\providecommand \@@endlink[0]{}%
\providecommand \url  [0]{\begingroup\@sanitize@url \@url }%
\providecommand \@url [1]{\endgroup\@href {#1}{\urlprefix }}%
\providecommand \urlprefix  [0]{URL }%
\providecommand \Eprint [0]{\href }%
\providecommand \doibase [0]{http://dx.doi.org/}%
\providecommand \selectlanguage [0]{\@gobble}%
\providecommand \bibinfo  [0]{\@secondoftwo}%
\providecommand \bibfield  [0]{\@secondoftwo}%
\providecommand \translation [1]{[#1]}%
\providecommand \BibitemOpen [0]{}%
\providecommand \bibitemStop [0]{}%
\providecommand \bibitemNoStop [0]{.\EOS\space}%
\providecommand \EOS [0]{\spacefactor3000\relax}%
\providecommand \BibitemShut  [1]{\csname bibitem#1\endcsname}%
\let\auto@bib@innerbib\@empty
%</preamble>
\bibitem [{\citenamefont {{Milton (editor)}}(2016)}]{Milton:2016:ETC}%
  \BibitemOpen
  \bibfield  {author} {\bibinfo {author} {\bibfnamefont {G.~W.}\ \bibnamefont
  {{Milton (editor)}}},\ }\href@noop {} {\emph {\bibinfo {title} {Extending the
  Theory of Composites to Other Areas of Science}}}\ (\bibinfo  {publisher}
  {Milton--Patton Publishers},\ \bibinfo {address} {P.O. Box 581077, Salt Lake
  City, UT 85148, USA},\ \bibinfo {year} {2016})\ pp.\ \bibinfo {pages} {xx +
  422}\BibitemShut {NoStop}%
\bibitem [{\citenamefont {Sharma}(2017)}]{Sharma:2017:BRE}%
  \BibitemOpen
  \bibfield  {author} {\bibinfo {author} {\bibfnamefont {P.}~\bibnamefont
  {Sharma}},\ }\href {\doibase http://dx.doi.org/10.1115/1.4035525} {\bibfield
  {journal} {\bibinfo  {journal} {Journal of Applied Mechanics}\ }\textbf
  {\bibinfo {volume} {84}},\ \bibinfo {pages} {036501} (\bibinfo {year}
  {2017})}\BibitemShut {NoStop}%
\bibitem [{\citenamefont {Grabovsky}(2017)}]{Grabovsky:2018:BRE}%
  \BibitemOpen
  \bibfield  {author} {\bibinfo {author} {\bibfnamefont {Y.}~\bibnamefont
  {Grabovsky}},\ }\href {\doibase http://dx.doi.org/10.1137/18N97456X}
  {\bibfield  {journal} {\bibinfo  {journal} {SIAM Review}\ }\textbf {\bibinfo
  {volume} {60}},\ \bibinfo {pages} {475} (\bibinfo {year} {2017})}\BibitemShut
  {NoStop}%
\bibitem [{\citenamefont {Cherkaev}(2000)}]{Cherkaev:2000:VMS}%
  \BibitemOpen
  \bibfield  {author} {\bibinfo {author} {\bibfnamefont {A.~V.}\ \bibnamefont
  {Cherkaev}},\ }\href {\doibase http://dx.doi.org/10.1007/978-1-4612-1188-4}
  {\emph {\bibinfo {title} {Variational Methods for Structural
  Optimization}}},\ \bibinfo {series} {Applied Mathematical Sciences}, Vol.\
  \bibinfo {volume} {140}\ (\bibinfo  {publisher} {Springer-Verlag},\ \bibinfo
  {address} {Berlin~/ Heidelberg~/ London~/ etc.},\ \bibinfo {year} {2000})\
  pp.\ \bibinfo {pages} {xxvi + 545}\BibitemShut {NoStop}%
\bibitem [{\citenamefont {Torquato}(2002)}]{Torquato:2002:RHM}%
  \BibitemOpen
  \bibfield  {author} {\bibinfo {author} {\bibfnamefont {S.}~\bibnamefont
  {Torquato}},\ }\href {\doibase http://dx.doi.org/10.1007/978-1-4757-6355-3}
  {\emph {\bibinfo {title} {Random Heterogeneous Materials: Microstructure and
  Macroscopic Properties}}},\ \bibinfo {series} {Interdisciplinary Applied
  Mathematics}, Vol.~\bibinfo {volume} {16}\ (\bibinfo  {publisher}
  {Springer-Verlag},\ \bibinfo {address} {Berlin, Germany~/ Heidelberg,
  Germany~/ London, UK~/ etc.},\ \bibinfo {year} {2002})\ pp.\ \bibinfo {pages}
  {xxi + 701}\BibitemShut {NoStop}%
\bibitem [{\citenamefont {Milton}(2002)}]{Milton:2002:TOC}%
  \BibitemOpen
  \bibfield  {author} {\bibinfo {author} {\bibfnamefont {G.~W.}\ \bibnamefont
  {Milton}},\ }\href {\doibase http://dx.doi.org/10.1017/CBO9780511613357}
  {\emph {\bibinfo {title} {The Theory of Composites}}},\ \bibinfo {series}
  {Cambridge Monographs on Applied and Computational Mathematics},
  Vol.~\bibinfo {volume} {6}\ (\bibinfo  {publisher} {Cambridge University
  Press},\ \bibinfo {address} {Cambridge, UK},\ \bibinfo {year} {2002})\ pp.\
  \bibinfo {pages} {xxviii + 719},\ \bibinfo {note} {series editors: P. G.
  Ciarlet, A. Iserles, Robert V. Kohn, and M. H. Wright.}\BibitemShut {Stop}%
\bibitem [{\citenamefont {Allaire}(2002)}]{Allaire:2002:SOH}%
  \BibitemOpen
  \bibfield  {author} {\bibinfo {author} {\bibfnamefont {G.}~\bibnamefont
  {Allaire}},\ }\href {\doibase http://dx.doi.org/10.1007/978-1-4684-9286-6}
  {\emph {\bibinfo {title} {Shape Optimization by the Homogenization
  Method}}},\ \bibinfo {series} {Applied Mathematical Sciences}, Vol.\ \bibinfo
  {volume} {146}\ (\bibinfo  {publisher} {Springer-Verlag},\ \bibinfo {address}
  {Berlin, Germany~/ Heidelberg, Germany~/ London, UK~/ etc.},\ \bibinfo {year}
  {2002})\ pp.\ \bibinfo {pages} {xv + 456}\BibitemShut {NoStop}%
\bibitem [{\citenamefont {Tartar}(2009)}]{Tartar:2009:GTH}%
  \BibitemOpen
  \bibfield  {author} {\bibinfo {author} {\bibfnamefont {L.}~\bibnamefont
  {Tartar}},\ }\href {\doibase http://dx.doi.org/10.1007/978-3-642-05195-1}
  {\emph {\bibinfo {title} {The General Theory of Homogenization: a
  Personalized Introduction}}},\ \bibinfo {series} {Lecture Notes of the Unione
  Matematica Italiana}, Vol.~\bibinfo {volume} {7}\ (\bibinfo  {publisher}
  {Springer-Verlag},\ \bibinfo {address} {Berlin, Germany~/ Heidelberg,
  Germany~/ London, UK~/ etc.},\ \bibinfo {year} {2009})\ pp.\ \bibinfo {pages}
  {xxii + 470}\BibitemShut {NoStop}%
\bibitem [{\citenamefont {Milton}\ and\ \citenamefont
  {Onofrei}(2018)}]{Milton:2018:ERG}%
  \BibitemOpen
  \bibfield  {author} {\bibinfo {author} {\bibfnamefont {G.~W.}\ \bibnamefont
  {Milton}}\ and\ \bibinfo {author} {\bibfnamefont {D.}~\bibnamefont
  {Onofrei}},\ }\href@noop {} {\bibfield  {journal} {\bibinfo  {journal}
  {Research in Mathematical Sciences}\ } (\bibinfo {year} {2018})},\ \bibinfo
  {note} {submitted. See arXiv:1712.03597 [math.AP]}\BibitemShut {NoStop}%
\bibitem [{\citenamefont {Grabovsky}(1998)}]{Grabovsky:1998:EREa}%
  \BibitemOpen
  \bibfield  {author} {\bibinfo {author} {\bibfnamefont {Y.}~\bibnamefont
  {Grabovsky}},\ }\href {\doibase http://dx.doi.org/10.1007/s002050050107}
  {\bibfield  {journal} {\bibinfo  {journal} {Archive for Rational Mechanics
  and Analysis}\ }\textbf {\bibinfo {volume} {143}},\ \bibinfo {pages} {309}
  (\bibinfo {year} {1998})}\BibitemShut {NoStop}%
\bibitem [{\citenamefont {Grabovsky}\ and\ \citenamefont
  {Sage}(1998)}]{Grabovsky:1998:EREb}%
  \BibitemOpen
  \bibfield  {author} {\bibinfo {author} {\bibfnamefont {Y.}~\bibnamefont
  {Grabovsky}}\ and\ \bibinfo {author} {\bibfnamefont {D.~S.}\ \bibnamefont
  {Sage}},\ }\href {\doibase http://dx.doi.org/10.1007/s002050050108}
  {\bibfield  {journal} {\bibinfo  {journal} {Archive for Rational Mechanics
  and Analysis}\ }\textbf {\bibinfo {volume} {143}},\ \bibinfo {pages} {331}
  (\bibinfo {year} {1998})}\BibitemShut {NoStop}%
\bibitem [{\citenamefont {Grabovsky}, \citenamefont {Milton},\ and\
  \citenamefont {Sage}(2000)}]{Grabovsky:2000:ERE}%
  \BibitemOpen
  \bibfield  {author} {\bibinfo {author} {\bibfnamefont {Y.}~\bibnamefont
  {Grabovsky}}, \bibinfo {author} {\bibfnamefont {G.~W.}\ \bibnamefont
  {Milton}}, \ and\ \bibinfo {author} {\bibfnamefont {D.~S.}\ \bibnamefont
  {Sage}},\ }\href {\doibase http://doi.org/d8k4vw} {\bibfield  {journal}
  {\bibinfo  {journal} {Communications on Pure and Applied Mathematics (New
  York)}\ }\textbf {\bibinfo {volume} {53}},\ \bibinfo {pages} {300} (\bibinfo
  {year} {2000})}\BibitemShut {NoStop}%
\bibitem [{\citenamefont {Grabovsky}(2016)}]{Grabovsky:2016:CMM}%
  \BibitemOpen
  \bibfield  {author} {\bibinfo {author} {\bibfnamefont {Y.}~\bibnamefont
  {Grabovsky}},\ }\href {\doibase http://dx.doi.org/10.1088/978-0-7503-1048-2}
  {\emph {\bibinfo {title} {Composite Materials: Mathematical Theory and Exact
  Relations}}}\ (\bibinfo  {publisher} {IOP Publishing},\ \bibinfo {address}
  {Bristol, UK},\ \bibinfo {year} {2016})\ pp.\ \bibinfo {pages} {xiv +
  209}\BibitemShut {NoStop}%
\bibitem [{\citenamefont {Horn}(1954)}]{Horn:1954:DSM}%
  \BibitemOpen
  \bibfield  {author} {\bibinfo {author} {\bibfnamefont {A.}~\bibnamefont
  {Horn}},\ }\href {\doibase http://dx.doi.org/10.2307/2372705} {\bibfield
  {journal} {\bibinfo  {journal} {American Journal of Mathematics}\ }\textbf
  {\bibinfo {volume} {76}},\ \bibinfo {pages} {620} (\bibinfo {year}
  {1954})}\BibitemShut {NoStop}%
\bibitem [{\citenamefont {Weyl}(1912)}]{Weyl:1912:AVE}%
  \BibitemOpen
  \bibfield  {author} {\bibinfo {author} {\bibfnamefont {H.}~\bibnamefont
  {Weyl}},\ }\href {\doibase http://dx.doi.org/10.1007/BF01456804} {\bibfield
  {journal} {\bibinfo  {journal} {Mathematische Annalen}\ }\textbf {\bibinfo
  {volume} {71}},\ \bibinfo {pages} {441} (\bibinfo {year} {1912})}\BibitemShut
  {NoStop}%
\bibitem [{\citenamefont {Lidski\u{i}}(1950)}]{Lidskii:1950:PVS}%
  \BibitemOpen
  \bibfield  {author} {\bibinfo {author} {\bibfnamefont {I.~M.}\ \bibnamefont
  {Lidski\u{i}}},\ }\href@noop {} {\bibfield  {journal} {\bibinfo  {journal}
  {Doklady Akademii Nauk SSSR}\ }\textbf {\bibinfo {volume} {75}},\ \bibinfo
  {pages} {769} (\bibinfo {year} {1950})},\ \bibinfo {note} {translated in
  National Bureau Standards Report 2248, February, 1953}\BibitemShut {NoStop}%
\bibitem [{\citenamefont {Wielandt}(1955)}]{Wielandt:1955:OSE}%
  \BibitemOpen
  \bibfield  {author} {\bibinfo {author} {\bibfnamefont {H.}~\bibnamefont
  {Wielandt}},\ }\href {\doibase
  http://dx.doi.org/10.1090/S0002-9939-1955-0067842-9} {\bibfield  {journal}
  {\bibinfo  {journal} {Proceedings of the American Mathematical Society}\
  }\textbf {\bibinfo {volume} {6}},\ \bibinfo {pages} {106} (\bibinfo {year}
  {1955})}\BibitemShut {NoStop}%
\bibitem [{\citenamefont {Horn}(1962)}]{Horn:1962:BPI}%
  \BibitemOpen
  \bibfield  {author} {\bibinfo {author} {\bibfnamefont {A.}~\bibnamefont
  {Horn}},\ }\href {https://projecteuclid.org/euclid.pjm/1103036720} {\bibfield
   {journal} {\bibinfo  {journal} {Pacific Journal of Mathematics}\ }\textbf
  {\bibinfo {volume} {12}},\ \bibinfo {pages} {225} (\bibinfo {year}
  {1962})}\BibitemShut {NoStop}%
\bibitem [{\citenamefont {Thompson}\ and\ \citenamefont
  {Garbanati}(1971)}]{Thompson:1971:ESH}%
  \BibitemOpen
  \bibfield  {author} {\bibinfo {author} {\bibfnamefont {R.~C.}\ \bibnamefont
  {Thompson}}\ and\ \bibinfo {author} {\bibfnamefont {L.~F.}\ \bibnamefont
  {Garbanati}},\ }\href {\doibase http://dx.doi.org/10.6028/jres.075B.007}
  {\bibfield  {journal} {\bibinfo  {journal} {Journal of Research of the
  National Bureau of Standards, Section B: Mathematical Sciences}\ }\textbf
  {\bibinfo {volume} {75B}},\ \bibinfo {pages} {115} (\bibinfo {year}
  {1971})}\BibitemShut {NoStop}%
\bibitem [{\citenamefont {Helmke}\ and\ \citenamefont
  {Rosenthal}(1995)}]{Helmke:1995:EIS}%
  \BibitemOpen
  \bibfield  {author} {\bibinfo {author} {\bibfnamefont {U.}~\bibnamefont
  {Helmke}}\ and\ \bibinfo {author} {\bibfnamefont {J.}~\bibnamefont
  {Rosenthal}},\ }\href {\doibase http://doi.org/10.1002/mana.19951710113}
  {\bibfield  {journal} {\bibinfo  {journal} {Mathematische Nachrichten}\
  }\textbf {\bibinfo {volume} {171}},\ \bibinfo {pages} {207} (\bibinfo {year}
  {1995})}\BibitemShut {NoStop}%
\bibitem [{\citenamefont {Klyachko}(1998)}]{Klyachko:1954:DSM}%
  \BibitemOpen
  \bibfield  {author} {\bibinfo {author} {\bibfnamefont {A.~A.}\ \bibnamefont
  {Klyachko}},\ }\href {\doibase http://dx.doi.org/10.1007/s000290050037}
  {\bibfield  {journal} {\bibinfo  {journal} {Selecta Mathematica}\ }\textbf
  {\bibinfo {volume} {4}},\ \bibinfo {pages} {419} (\bibinfo {year}
  {1998})}\BibitemShut {NoStop}%
\bibitem [{\citenamefont {Fulton}(2000)}]{Fulton:2000:EIF}%
  \BibitemOpen
  \bibfield  {author} {\bibinfo {author} {\bibfnamefont {W.}~\bibnamefont
  {Fulton}},\ }\href {\doibase https://doi.org/10.1090/S0273-0979-00-00865-X}
  {\bibfield  {journal} {\bibinfo  {journal} {Bulletin of the American
  Mathematical Society}\ }\textbf {\bibinfo {volume} {37}},\ \bibinfo {pages}
  {209} (\bibinfo {year} {2000})}\BibitemShut {NoStop}%
\bibitem [{\citenamefont {Tonelli}(1920)}]{Tonelli:1920:SCV}%
  \BibitemOpen
  \bibfield  {author} {\bibinfo {author} {\bibfnamefont {L.}~\bibnamefont
  {Tonelli}},\ }\href@noop {} {\bibfield  {journal} {\bibinfo  {journal}
  {Rendiconti Circolo Matematico di Palermo}\ }\textbf {\bibinfo {volume}
  {44}},\ \bibinfo {pages} {167} (\bibinfo {year} {1920})}\BibitemShut
  {NoStop}%
\bibitem [{\citenamefont {Terpstra}(1938)}]{Terpstra:1938:DBF}%
  \BibitemOpen
  \bibfield  {author} {\bibinfo {author} {\bibfnamefont {F.~J.}\ \bibnamefont
  {Terpstra}},\ }\href {\doibase http://dx.doi.org/10.1007/BF01597353}
  {\bibfield  {journal} {\bibinfo  {journal} {Mathematische Annalen}\ }\textbf
  {\bibinfo {volume} {116}},\ \bibinfo {pages} {166} (\bibinfo {year}
  {1938})}\BibitemShut {NoStop}%
\bibitem [{\citenamefont {Morrey}(1952)}]{Morrey:1952:QSM}%
  \BibitemOpen
  \bibfield  {author} {\bibinfo {author} {\bibfnamefont {C.~B.}\ \bibnamefont
  {Morrey}},\ }\href {\doibase http://dx.doi.org/10.2140/pjm.1952.2.25}
  {\bibfield  {journal} {\bibinfo  {journal} {Pacific Journal of Mathematics}\
  }\textbf {\bibinfo {volume} {2}},\ \bibinfo {pages} {25} (\bibinfo {year}
  {1952})}\BibitemShut {NoStop}%
\bibitem [{\citenamefont {Meyers}(1965)}]{Meyers:1965:ASA}%
  \BibitemOpen
  \bibfield  {author} {\bibinfo {author} {\bibfnamefont {N.~G.}\ \bibnamefont
  {Meyers}},\ }\href {\doibase http://dx.doi.org/10.2307/1994235} {\bibfield
  {journal} {\bibinfo  {journal} {Transactions of the American Mathematical
  Society}\ }\textbf {\bibinfo {volume} {119}},\ \bibinfo {pages} {125}
  (\bibinfo {year} {1965})}\BibitemShut {NoStop}%
\bibitem [{\citenamefont {Morrey}(1966)}]{Morrey:1966:MIC}%
  \BibitemOpen
  \bibfield  {author} {\bibinfo {author} {\bibfnamefont {C.~B.}\ \bibnamefont
  {Morrey}},\ }\enquote {\bibinfo {title} {Multiple integrals in the calculus
  of variations},}\ \ (\bibinfo  {publisher} {Springer-Verlag},\ \bibinfo
  {address} {Berlin~/ Heidelberg~/ London~/ etc.},\ \bibinfo {year} {1966})\
  p.\ \bibinfo {pages} {122}\BibitemShut {NoStop}%
\bibitem [{\citenamefont {Murat}(1978)}]{Murat:1978:CPC}%
  \BibitemOpen
  \bibfield  {author} {\bibinfo {author} {\bibfnamefont {F.}~\bibnamefont
  {Murat}},\ }\href {http://www.numdam.org/item?id=ASNSP_1978_4_5_3_489_0}
  {\bibfield  {journal} {\bibinfo  {journal} {Annali della Scuola normale
  superiore di Pisa, Classe di scienze. Serie IV}\ }\textbf {\bibinfo {volume}
  {5}},\ \bibinfo {pages} {489} (\bibinfo {year} {1978})}\BibitemShut {NoStop}%
\bibitem [{\citenamefont {Tartar}(1979{\natexlab{a}})}]{Tartar:1979:CCA}%
  \BibitemOpen
  \bibfield  {author} {\bibinfo {author} {\bibfnamefont {L.}~\bibnamefont
  {Tartar}},\ }in\ \href@noop {} {\emph {\bibinfo {booktitle} {Nonlinear
  Analysis and Mechanics, Heriot--Watt Symposium, Volume IV}}},\ \bibinfo
  {series} {Research Notes in Mathematics}, Vol.~\bibinfo {volume} {39},\
  \bibinfo {editor} {edited by\ \bibinfo {editor} {\bibfnamefont {R.~J.}\
  \bibnamefont {Knops}}}\ (\bibinfo  {publisher} {Pitman Publishing Ltd.},\
  \bibinfo {address} {London},\ \bibinfo {year} {1979})\ pp.\ \bibinfo {pages}
  {136--212}\BibitemShut {NoStop}%
\bibitem [{\citenamefont {Tartar}(1979{\natexlab{b}})}]{Tartar:1979:ECH}%
  \BibitemOpen
  \bibfield  {author} {\bibinfo {author} {\bibfnamefont {L.}~\bibnamefont
  {Tartar}},\ }in\ \href@noop {} {\emph {\bibinfo {booktitle} {{Computing
  Methods in Applied Sciences and Engineering: Third International Symposium,
  Versailles, France, December 5--9, 1977}}}},\ \bibinfo {series} {Lecture
  Notes in Mathematics}, Vol.\ \bibinfo {volume} {704},\ \bibinfo {editor}
  {edited by\ \bibinfo {editor} {\bibfnamefont {R.}~\bibnamefont {Glowinski}}\
  and\ \bibinfo {editor} {\bibfnamefont {J.-L.}\ \bibnamefont {Lions}}}\
  (\bibinfo  {publisher} {Springer-Verlag},\ \bibinfo {address} {Berlin~/
  Heidelberg~/ London~/ etc.},\ \bibinfo {year} {1979})\ pp.\ \bibinfo {pages}
  {364--373},\ \bibinfo {note} {english translation in {Topics in
  the Mathematical Modelling of Composite Materials}, pp. 9--20, ed. by A.
  Cherkaev and R. Kohn. ISBN 0-8176-3662-5.}\BibitemShut {Stop}%
\bibitem [{\citenamefont {Ball}, \citenamefont {Currie},\ and\ \citenamefont
  {Olver}(1981)}]{Ball:1981:NLW}%
  \BibitemOpen
  \bibfield  {author} {\bibinfo {author} {\bibfnamefont {J.~M.}\ \bibnamefont
  {Ball}}, \bibinfo {author} {\bibfnamefont {J.~C.}\ \bibnamefont {Currie}}, \
  and\ \bibinfo {author} {\bibfnamefont {P.~J.}\ \bibnamefont {Olver}},\ }\href
  {\doibase http://dx.doi.org/10.1016/0022-1236(81)90085-9} {\bibfield
  {journal} {\bibinfo  {journal} {Journal of Functional Analysis}\ }\textbf
  {\bibinfo {volume} {41}},\ \bibinfo {pages} {135} (\bibinfo {year}
  {1981})}\BibitemShut {NoStop}%
\bibitem [{\citenamefont {Murat}(1981)}]{Murat:1981:CPC}%
  \BibitemOpen
  \bibfield  {author} {\bibinfo {author} {\bibfnamefont {F.}~\bibnamefont
  {Murat}},\ }\href {http://www.numdam.org/item?id=ASNSP_1981_4_8_1_69_0}
  {\bibfield  {journal} {\bibinfo  {journal} {Annali della Scuola normale
  superiore di Pisa, Classe di scienze. Serie IV}\ }\textbf {\bibinfo {volume}
  {8}},\ \bibinfo {pages} {69} (\bibinfo {year} {1981})}\BibitemShut {NoStop}%
\bibitem [{\citenamefont {Murat}\ and\ \citenamefont
  {Tartar}(1985)}]{Murat:1985:CVH}%
  \BibitemOpen
  \bibfield  {author} {\bibinfo {author} {\bibfnamefont {F.}~\bibnamefont
  {Murat}}\ and\ \bibinfo {author} {\bibfnamefont {L.}~\bibnamefont {Tartar}},\
  }in\ \href@noop {} {\emph {\bibinfo {booktitle} {Les m{\'e}thodes de
  l'homog{\'e}n{\'e}isation: th{\'e}orie et applications en physique}}},\
  \bibinfo {series} {Collection de la Direction des {\'e}tudes et recherches
  d'{\'E}lectricit{\'e} de France}, Vol.~\bibinfo {volume} {57}\ (\bibinfo
  {publisher} {Eyrolles},\ \bibinfo {address} {Paris},\ \bibinfo {year}
  {1985})\ pp.\ \bibinfo {pages} {319--369},\ \bibinfo {note} {english
  translation in {Topics in the Mathematical Modelling of Composite
  Materials}, pp. 139--173, ed. by A. Cherkaev and R. Kohn, ISBN
  0-8176-3662-5.}\BibitemShut {Stop}%
\bibitem [{\citenamefont {Tartar}(1985)}]{Tartar:1985:EFC}%
  \BibitemOpen
  \bibfield  {author} {\bibinfo {author} {\bibfnamefont {L.}~\bibnamefont
  {Tartar}},\ }in\ \href@noop {} {\emph {\bibinfo {booktitle} {Ennio de Giorgi
  Colloquium: Papers Presented at a Colloquium Held at the {H. Poincar{\'e}
  Institute} in November 1983}}},\ \bibinfo {series} {Pitman Research Notes in
  Mathematics}, Vol.\ \bibinfo {volume} {125},\ \bibinfo {editor} {edited by\
  \bibinfo {editor} {\bibfnamefont {P.}~\bibnamefont {Kr{\'e}e}}}\ (\bibinfo
  {publisher} {Pitman Publishing Ltd.},\ \bibinfo {address} {London},\ \bibinfo
  {year} {1985})\ pp.\ \bibinfo {pages} {168--187}\BibitemShut {NoStop}%
\bibitem [{\citenamefont {Ben\v{e}sov\'{a}}\ and\ \citenamefont
  {Kru\v{z}\'{i}k}(2017)}]{Benesova:2017:WLS}%
  \BibitemOpen
  \bibfield  {author} {\bibinfo {author} {\bibfnamefont {B.}~\bibnamefont
  {Ben\v{e}sov\'{a}}}\ and\ \bibinfo {author} {\bibfnamefont {M.}~\bibnamefont
  {Kru\v{z}\'{i}k}},\ }\href {\doibase http://dx.doi.org/10.1137/16M1060947}
  {\bibfield  {journal} {\bibinfo  {journal} {SIAM Review}\ }\textbf {\bibinfo
  {volume} {59}},\ \bibinfo {pages} {703} (\bibinfo {year} {2017})}\BibitemShut
  {NoStop}%
\bibitem [{\citenamefont {Milton}(2013)}]{Milton:2013:SIG}%
  \BibitemOpen
  \bibfield  {author} {\bibinfo {author} {\bibfnamefont {G.~W.}\ \bibnamefont
  {Milton}},\ }\href {\doibase http://dx.doi.org/10.1098/rspa.2013.0075}
  {\bibfield  {journal} {\bibinfo  {journal} {Proceedings of the Royal Society
  A: Mathematical, Physical, \& Engineering Sciences}\ }\textbf {\bibinfo
  {volume} {469}},\ \bibinfo {pages} {20130075} (\bibinfo {year} {2013})},\
  \bibinfo {note} {see addendum \cite{Milton:2015:ATS}.}\BibitemShut {Stop}%
\bibitem [{\citenamefont {Milton}(2015)}]{Milton:2015:ATS}%
  \BibitemOpen
  \bibfield  {author} {\bibinfo {author} {\bibfnamefont {G.~W.}\ \bibnamefont
  {Milton}},\ }\href {\doibase http://dx.doi.org/10.1098/rspa.2014.0886}
  {\bibfield  {journal} {\bibinfo  {journal} {Proceedings of the Royal Society
  A: Mathematical, Physical, \& Engineering Sciences}\ }\textbf {\bibinfo
  {volume} {471}},\ \bibinfo {pages} {20140886} (\bibinfo {year} {2015})},\
  \bibinfo {note} {see \cite{Milton:2013:SIG}.}\BibitemShut {Stop}%
\bibitem [{\citenamefont {Buliga}(2008)}]{Buliga:208:FAM}%
  \BibitemOpen
  \bibfield  {author} {\bibinfo {author} {\bibfnamefont {M.}~\bibnamefont
  {Buliga}},\ }\href {\doibase http://dx.doi.org/10.1016/j.laa.2008.04.023}
  {\bibfield  {journal} {\bibinfo  {journal} {Linear Algebra and its
  Applications}\ }\textbf {\bibinfo {volume} {429}},\ \bibinfo {pages} {1528}
  (\bibinfo {year} {2008})}\BibitemShut {NoStop}%
\bibitem [{\citenamefont {Strang}(1988)}]{Strang:1988:FEE}%
  \BibitemOpen
  \bibfield  {author} {\bibinfo {author} {\bibfnamefont {G.}~\bibnamefont
  {Strang}},\ }\href {\doibase http://dx.doi.org/10.1137/1030048} {\bibfield
  {journal} {\bibinfo  {journal} {SIAM Review}\ }\textbf {\bibinfo {volume}
  {30}},\ \bibinfo {pages} {283} (\bibinfo {year} {1988})}\BibitemShut
  {NoStop}%
\bibitem [{\citenamefont {Milton}, \citenamefont {Seppecher},\ and\
  \citenamefont {Bouchitt{\'e}}(2009)}]{Milton:2009:MVP}%
  \BibitemOpen
  \bibfield  {author} {\bibinfo {author} {\bibfnamefont {G.~W.}\ \bibnamefont
  {Milton}}, \bibinfo {author} {\bibfnamefont {P.}~\bibnamefont {Seppecher}}, \
  and\ \bibinfo {author} {\bibfnamefont {G.}~\bibnamefont {Bouchitt{\'e}}},\
  }\href {\doibase http://dx.doi.org/10.1098/rspa.2008.0195} {\bibfield
  {journal} {\bibinfo  {journal} {Proceedings of the Royal Society A:
  Mathematical, Physical, \& Engineering Sciences}\ }\textbf {\bibinfo {volume}
  {465}},\ \bibinfo {pages} {367} (\bibinfo {year} {2009})}\BibitemShut
  {NoStop}%
\bibitem [{\citenamefont {Schoenberg}\ and\ \citenamefont
  {Sen}(1983)}]{Schoenberg:1983:PPS}%
  \BibitemOpen
  \bibfield  {author} {\bibinfo {author} {\bibfnamefont {M.}~\bibnamefont
  {Schoenberg}}\ and\ \bibinfo {author} {\bibfnamefont {P.~N.}\ \bibnamefont
  {Sen}},\ }\href {\doibase http://dx.doi.org/10.1121/1.388724} {\bibfield
  {journal} {\bibinfo  {journal} {Journal of the Acoustical Society of
  America}\ }\textbf {\bibinfo {volume} {73}},\ \bibinfo {pages} {61} (\bibinfo
  {year} {1983})}\BibitemShut {NoStop}%
\bibitem [{\citenamefont {Willis}(1985)}]{Willis:1985:NID}%
  \BibitemOpen
  \bibfield  {author} {\bibinfo {author} {\bibfnamefont {J.~R.}\ \bibnamefont
  {Willis}},\ }\href {\doibase http://dx.doi.org/10.1016/0020-7683(85)90084-8}
  {\bibfield  {journal} {\bibinfo  {journal} {International Journal of Solids
  and Structures}\ }\textbf {\bibinfo {volume} {21}},\ \bibinfo {pages} {805}
  (\bibinfo {year} {1985})}\BibitemShut {NoStop}%
\bibitem [{\citenamefont {Milton}, \citenamefont {Briane},\ and\ \citenamefont
  {Willis}(2006)}]{Milton:2006:CEM}%
  \BibitemOpen
  \bibfield  {author} {\bibinfo {author} {\bibfnamefont {G.~W.}\ \bibnamefont
  {Milton}}, \bibinfo {author} {\bibfnamefont {M.}~\bibnamefont {Briane}}, \
  and\ \bibinfo {author} {\bibfnamefont {J.~R.}\ \bibnamefont {Willis}},\
  }\href {\doibase http://dx.doi.org/10.1088/1367-2630/8/10/248} {\bibfield
  {journal} {\bibinfo  {journal} {New Journal of Physics}\ }\textbf {\bibinfo
  {volume} {8}},\ \bibinfo {pages} {248} (\bibinfo {year} {2006})}\BibitemShut
  {NoStop}%
\bibitem [{\citenamefont {Milton}\ and\ \citenamefont
  {Willis}(2007)}]{Milton:2007:MNS}%
  \BibitemOpen
  \bibfield  {author} {\bibinfo {author} {\bibfnamefont {G.~W.}\ \bibnamefont
  {Milton}}\ and\ \bibinfo {author} {\bibfnamefont {J.~R.}\ \bibnamefont
  {Willis}},\ }\href {\doibase http://dx.doi.org/10.1098/rspa.2006.1795}
  {\bibfield  {journal} {\bibinfo  {journal} {Proceedings of the Royal Society
  A: Mathematical, Physical, \& Engineering Sciences}\ }\textbf {\bibinfo
  {volume} {463}},\ \bibinfo {pages} {855} (\bibinfo {year}
  {2007})}\BibitemShut {NoStop}%
\bibitem [{\citenamefont {Cherkaev}\ and\ \citenamefont
  {Gibiansky}(1994)}]{Cherkaev:1994:VPC}%
  \BibitemOpen
  \bibfield  {author} {\bibinfo {author} {\bibfnamefont {A.~V.}\ \bibnamefont
  {Cherkaev}}\ and\ \bibinfo {author} {\bibfnamefont {L.~V.}\ \bibnamefont
  {Gibiansky}},\ }\href {\doibase http://dx.doi.org/10.1063/1.530782}
  {\bibfield  {journal} {\bibinfo  {journal} {Journal of Mathematical Physics}\
  }\textbf {\bibinfo {volume} {35}},\ \bibinfo {pages} {127} (\bibinfo {year}
  {1994})}\BibitemShut {NoStop}%
\bibitem [{\citenamefont {Milton}\ and\ \citenamefont
  {Willis}(2010)}]{Milton:2010:MVP}%
  \BibitemOpen
  \bibfield  {author} {\bibinfo {author} {\bibfnamefont {G.~W.}\ \bibnamefont
  {Milton}}\ and\ \bibinfo {author} {\bibfnamefont {J.~R.}\ \bibnamefont
  {Willis}},\ }\href {\doibase http://dx.doi.org/10.1098/rspa.2010.0006}
  {\bibfield  {journal} {\bibinfo  {journal} {Proceedings of the Royal Society
  A: Mathematical, Physical, \& Engineering Sciences}\ }\textbf {\bibinfo
  {volume} {466}},\ \bibinfo {pages} {3013} (\bibinfo {year}
  {2010})}\BibitemShut {NoStop}%
\bibitem [{\citenamefont {Parr}\ and\ \citenamefont
  {Weitao}(1994)}]{Parr:1994:DFT}%
  \BibitemOpen
  \bibfield  {author} {\bibinfo {author} {\bibfnamefont {R.~G.}\ \bibnamefont
  {Parr}}\ and\ \bibinfo {author} {\bibfnamefont {Y.}~\bibnamefont {Weitao}},\
  }\href {http://catdir.loc.gov/catdir/enhancements/fy0639/88025157-d.html;
  http://catdir.loc.gov/catdir/enhancements/fy0639/88025157-t.html} {\emph
  {\bibinfo {title} {Density-Functional Theory of Atoms and Molecules}}},\
  \bibinfo {series} {International Series of Monographs on Chemistry},
  Vol.~\bibinfo {volume} {16}\ (\bibinfo  {publisher} {Oxford University
  Press},\ \bibinfo {address} {Oxford, UK},\ \bibinfo {year} {1994})\ pp.\
  \bibinfo {pages} {ix + 333}\BibitemShut {NoStop}%
\bibitem [{\citenamefont {Milton}(2017)}]{Milton:2017:BCP}%
  \BibitemOpen
  \bibfield  {author} {\bibinfo {author} {\bibfnamefont {G.~W.}\ \bibnamefont
  {Milton}},\ }\href {\doibase https://doi.org/10.1103/PhysRevB.96.104206}
  {\bibfield  {journal} {\bibinfo  {journal} {Physical Review B: Condensed
  Matter and Materials Physics}\ }\textbf {\bibinfo {volume} {96}},\ \bibinfo
  {pages} {104206} (\bibinfo {year} {2017})}\BibitemShut {NoStop}%
\bibitem [{\citenamefont {Ilyin}(2010)}]{Ilyin:2006:LBS}%
  \BibitemOpen
  \bibfield  {author} {\bibinfo {author} {\bibfnamefont {A.~A.}\ \bibnamefont
  {Ilyin}},\ }\href {\doibase http://dx.doi.org/10.3934/dcds.2010.28.131}
  {\bibfield  {journal} {\bibinfo  {journal} {Discrete \& Continuous Dynamical
  Systems -- A}\ }\textbf {\bibinfo {volume} {28}},\ \bibinfo {pages} {131}
  (\bibinfo {year} {2010})}\BibitemShut {NoStop}%
\bibitem [{\citenamefont {Rudin}(1987)}]{Rudin:1987:RCA}%
  \BibitemOpen
  \bibfield  {author} {\bibinfo {author} {\bibfnamefont {W.}~\bibnamefont
  {Rudin}},\ }\href@noop {} {\emph {\bibinfo {title} {Real and Complex
  Analysis}}}\ (\bibinfo  {publisher} {Mc{\-}Graw-Hill},\ \bibinfo {address}
  {New York},\ \bibinfo {year} {1987})\ pp.\ \bibinfo {pages} {xiv +
  416}\BibitemShut {NoStop}%
\bibitem [{\citenamefont {Bergman}(1978)}]{Bergman:1978:DCC}%
  \BibitemOpen
  \bibfield  {author} {\bibinfo {author} {\bibfnamefont {D.~J.}\ \bibnamefont
  {Bergman}},\ }\href {\doibase http://dx.doi.org/10.1016/0370-1573(78)90009-1}
  {\bibfield  {journal} {\bibinfo  {journal} {Physics Reports}\ }\textbf
  {\bibinfo {volume} {43}},\ \bibinfo {pages} {377} (\bibinfo {year}
  {1978})}\BibitemShut {NoStop}%
\bibitem [{\citenamefont {Milton}(1979)}]{Milton:1979:TST}%
  \BibitemOpen
  \bibfield  {author} {\bibinfo {author} {\bibfnamefont {G.~W.}\ \bibnamefont
  {Milton}},\ }\href {\doibase http://dx.doi.org/10.13140/RG.2.1.2184.8482}
  {\enquote {\bibinfo {title} {Theoretical studies of the transport properties
  of inhomogeneous media},}\ }\bibinfo {type} {Unpublished report}\ \bibinfo
  {number} {TP/79/1}\ (\bibinfo  {institution} {University of Sydney},\
  \bibinfo {address} {Sydney, Australia},\ \bibinfo {year} {1979})\BibitemShut
  {NoStop}%
\bibitem [{\citenamefont {Milton}(1981{\natexlab{a}})}]{Milton:1981:BCP}%
  \BibitemOpen
  \bibfield  {author} {\bibinfo {author} {\bibfnamefont {G.~W.}\ \bibnamefont
  {Milton}},\ }\href {\doibase http://dx.doi.org/10.1063/1.329385} {\bibfield
  {journal} {\bibinfo  {journal} {Journal of Applied Physics}\ }\textbf
  {\bibinfo {volume} {52}},\ \bibinfo {pages} {5286} (\bibinfo {year}
  {1981}{\natexlab{a}})}\BibitemShut {NoStop}%
\bibitem [{\citenamefont {Milton}(1981{\natexlab{b}})}]{Milton:1981:BTO}%
  \BibitemOpen
  \bibfield  {author} {\bibinfo {author} {\bibfnamefont {G.~W.}\ \bibnamefont
  {Milton}},\ }\href {\doibase http://dx.doi.org/10.1063/1.329386} {\bibfield
  {journal} {\bibinfo  {journal} {Journal of Applied Physics}\ }\textbf
  {\bibinfo {volume} {52}},\ \bibinfo {pages} {5294} (\bibinfo {year}
  {1981}{\natexlab{b}})}\BibitemShut {NoStop}%
\bibitem [{\citenamefont {Bergman}(1980)}]{Bergman:1980:ESM}%
  \BibitemOpen
  \bibfield  {author} {\bibinfo {author} {\bibfnamefont {D.~J.}\ \bibnamefont
  {Bergman}},\ }\href {\doibase http://dx.doi.org/10.1103/PhysRevLett.44.1285}
  {\bibfield  {journal} {\bibinfo  {journal} {Physical Review Letters}\
  }\textbf {\bibinfo {volume} {44}},\ \bibinfo {pages} {1285} (\bibinfo {year}
  {1980})}\BibitemShut {NoStop}%
\bibitem [{\citenamefont {Bergman}(1982)}]{Bergman:1982:RBC}%
  \BibitemOpen
  \bibfield  {author} {\bibinfo {author} {\bibfnamefont {D.~J.}\ \bibnamefont
  {Bergman}},\ }\href {\doibase http://dx.doi.org/10.1016/0003-4916(82)90176-2}
  {\bibfield  {journal} {\bibinfo  {journal} {Annals of Physics}\ }\textbf
  {\bibinfo {volume} {138}},\ \bibinfo {pages} {78} (\bibinfo {year}
  {1982})}\BibitemShut {NoStop}%
\bibitem [{\citenamefont {Dell'Antonio}, \citenamefont {Figari},\ and\
  \citenamefont {Orlandi}(1986)}]{DellAntonio:1986:ATO}%
  \BibitemOpen
  \bibfield  {author} {\bibinfo {author} {\bibfnamefont {G.~F.}\ \bibnamefont
  {Dell'Antonio}}, \bibinfo {author} {\bibfnamefont {R.}~\bibnamefont
  {Figari}}, \ and\ \bibinfo {author} {\bibfnamefont {E.}~\bibnamefont
  {Orlandi}},\ }\href {http://eudml.org/doc/76310} {\bibfield  {journal}
  {\bibinfo  {journal} {Annales de l'institut {Henri Poincar{\'e} (A)} Physique
  th{\'e}orique}\ }\textbf {\bibinfo {volume} {44}},\ \bibinfo {pages} {1}
  (\bibinfo {year} {1986})}\BibitemShut {NoStop}%
\bibitem [{\citenamefont {Golden}\ and\ \citenamefont
  {Papanicolaou}(1983)}]{Golden:1983:BEP}%
  \BibitemOpen
  \bibfield  {author} {\bibinfo {author} {\bibfnamefont {K.~M.}\ \bibnamefont
  {Golden}}\ and\ \bibinfo {author} {\bibfnamefont {G.~C.}\ \bibnamefont
  {Papanicolaou}},\ }\href {\doibase http://dx.doi.org/10.1007/BF01216179}
  {\bibfield  {journal} {\bibinfo  {journal} {Communications in Mathematical
  Physics}\ }\textbf {\bibinfo {volume} {90}},\ \bibinfo {pages} {473}
  (\bibinfo {year} {1983})}\BibitemShut {NoStop}%
\bibitem [{\citenamefont {Golden}\ and\ \citenamefont
  {Papanicolaou}(1985)}]{Golden:1985:BEP}%
  \BibitemOpen
  \bibfield  {author} {\bibinfo {author} {\bibfnamefont {K.~M.}\ \bibnamefont
  {Golden}}\ and\ \bibinfo {author} {\bibfnamefont {G.~C.}\ \bibnamefont
  {Papanicolaou}},\ }\href {\doibase http://dx.doi.org/10.1007/BF01009895}
  {\bibfield  {journal} {\bibinfo  {journal} {Journal of Statistical Physics}\
  }\textbf {\bibinfo {volume} {40}},\ \bibinfo {pages} {655} (\bibinfo {year}
  {1985})}\BibitemShut {NoStop}%
\bibitem [{\citenamefont {Clark}\ and\ \citenamefont
  {Milton}(1994)}]{Clark:1994:MEC}%
  \BibitemOpen
  \bibfield  {author} {\bibinfo {author} {\bibfnamefont {K.~E.}\ \bibnamefont
  {Clark}}\ and\ \bibinfo {author} {\bibfnamefont {G.~W.}\ \bibnamefont
  {Milton}},\ }\href {\doibase http://dx.doi.org/10.1017/S030821050002864X}
  {\bibfield  {journal} {\bibinfo  {journal} {Proceedings of the Royal Society
  of Edinburgh}\ }\textbf {\bibinfo {volume} {124A}},\ \bibinfo {pages} {757}
  (\bibinfo {year} {1994})}\BibitemShut {NoStop}%
\bibitem [{\citenamefont {Clark}\ and\ \citenamefont
  {Milton}(1995)}]{Clark:1995:OBC}%
  \BibitemOpen
  \bibfield  {author} {\bibinfo {author} {\bibfnamefont {K.~E.}\ \bibnamefont
  {Clark}}\ and\ \bibinfo {author} {\bibfnamefont {G.~W.}\ \bibnamefont
  {Milton}},\ }\href {\doibase http://dx.doi.org/10.1098/rspa.1995.0011}
  {\bibfield  {journal} {\bibinfo  {journal} {Proceedings of the Royal Society
  of London. Series A, Mathematical and Physical Sciences}\ }\textbf {\bibinfo
  {volume} {448}},\ \bibinfo {pages} {161} (\bibinfo {year}
  {1995})}\BibitemShut {NoStop}%
\bibitem [{\citenamefont {Clark}(1997)}]{Clark:1997:CFR}%
  \BibitemOpen
  \bibfield  {author} {\bibinfo {author} {\bibfnamefont {K.~E.}\ \bibnamefont
  {Clark}},\ }\href {\doibase http://dx.doi.org/10.1063/1.532141} {\bibfield
  {journal} {\bibinfo  {journal} {Journal of Mathematical Physics}\ }\textbf
  {\bibinfo {volume} {38}},\ \bibinfo {pages} {4528} (\bibinfo {year}
  {1997})}\BibitemShut {NoStop}%
\bibitem [{\citenamefont {Avellaneda}\ \emph {et~al.}(1988)\citenamefont
  {Avellaneda}, \citenamefont {Cherkaev}, \citenamefont {Lurie},\ and\
  \citenamefont {Milton}}]{Avellaneda:1988:ECP}%
  \BibitemOpen
  \bibfield  {author} {\bibinfo {author} {\bibfnamefont {M.}~\bibnamefont
  {Avellaneda}}, \bibinfo {author} {\bibfnamefont {A.~V.}\ \bibnamefont
  {Cherkaev}}, \bibinfo {author} {\bibfnamefont {K.~A.}\ \bibnamefont {Lurie}},
  \ and\ \bibinfo {author} {\bibfnamefont {G.~W.}\ \bibnamefont {Milton}},\
  }\href {\doibase http://dx.doi.org/10.1063/1.340445} {\bibfield  {journal}
  {\bibinfo  {journal} {Journal of Applied Physics}\ }\textbf {\bibinfo
  {volume} {63}},\ \bibinfo {pages} {4989} (\bibinfo {year}
  {1988})}\BibitemShut {NoStop}%
\bibitem [{\citenamefont {Olver}\ and\ \citenamefont
  {Sivaloganathan}(1988)}]{Olver:1988:SNL}%
  \BibitemOpen
  \bibfield  {author} {\bibinfo {author} {\bibfnamefont {P.~J.}\ \bibnamefont
  {Olver}}\ and\ \bibinfo {author} {\bibfnamefont {J.}~\bibnamefont
  {Sivaloganathan}},\ }\href {\doibase
  http://dx.doi.org/10.1088/0951-7715/1/2/005} {\bibfield  {journal} {\bibinfo
  {journal} {Nonlinearity (Bristol)}\ }\textbf {\bibinfo {volume} {1}},\
  \bibinfo {pages} {389} (\bibinfo {year} {1988})}\BibitemShut {NoStop}%
\bibitem [{\citenamefont {Kang}\ and\ \citenamefont
  {Milton}(2013)}]{Kang:2013:BVF3d}%
  \BibitemOpen
  \bibfield  {author} {\bibinfo {author} {\bibfnamefont {H.}~\bibnamefont
  {Kang}}\ and\ \bibinfo {author} {\bibfnamefont {G.~W.}\ \bibnamefont
  {Milton}},\ }\href {\doibase http://dx.doi.org/10.1137/120879713} {\bibfield
  {journal} {\bibinfo  {journal} {SIAM Journal on Applied Mathematics}\
  }\textbf {\bibinfo {volume} {73}},\ \bibinfo {pages} {475} (\bibinfo {year}
  {2013})}\BibitemShut {NoStop}%
\bibitem [{\citenamefont {Kohn}\ and\ \citenamefont
  {Lipton}(1988)}]{Kohn:1988:OBE}%
  \BibitemOpen
  \bibfield  {author} {\bibinfo {author} {\bibfnamefont {R.~V.}\ \bibnamefont
  {Kohn}}\ and\ \bibinfo {author} {\bibfnamefont {R.}~\bibnamefont {Lipton}},\
  }\href {\doibase http://dx.doi.org/10.1007/BF00251534} {\bibfield  {journal}
  {\bibinfo  {journal} {Archive for Rational Mechanics and Analysis}\ }\textbf
  {\bibinfo {volume} {102}},\ \bibinfo {pages} {331} (\bibinfo {year}
  {1988})}\BibitemShut {NoStop}%
\bibitem [{\citenamefont {Milton}(1990)}]{Milton:1990:CSP}%
  \BibitemOpen
  \bibfield  {author} {\bibinfo {author} {\bibfnamefont {G.~W.}\ \bibnamefont
  {Milton}},\ }\href {\doibase http://dx.doi.org/10.1002/cpa.3160430104}
  {\bibfield  {journal} {\bibinfo  {journal} {Communications on Pure and
  Applied Mathematics (New York)}\ }\textbf {\bibinfo {volume} {43}},\ \bibinfo
  {pages} {63} (\bibinfo {year} {1990})}\BibitemShut {NoStop}%
\bibitem [{\citenamefont {Harutyunyan}\ and\ \citenamefont
  {Milton}(2015)}]{Harutyunyan:2015:EEE}%
  \BibitemOpen
  \bibfield  {author} {\bibinfo {author} {\bibfnamefont {D.}~\bibnamefont
  {Harutyunyan}}\ and\ \bibinfo {author} {\bibfnamefont {G.~W.}\ \bibnamefont
  {Milton}},\ }\href {\doibase http://dx.doi.org/10.1007/s00526-015-0836-z}
  {\bibfield  {journal} {\bibinfo  {journal} {Calculus of Variations and
  Partial Differential Equations}\ }\textbf {\bibinfo {volume} {54}},\ \bibinfo
  {pages} {1575} (\bibinfo {year} {2015})},\ \bibinfo {note} {see also
  arXiv:1403.3718 [math.AP].}\BibitemShut {Stop}%
\bibitem [{\citenamefont {Harutyunyan}\ and\ \citenamefont
  {Milton}(2017{\natexlab{a}})}]{Harutyunyan:2015:REE}%
  \BibitemOpen
  \bibfield  {author} {\bibinfo {author} {\bibfnamefont {D.}~\bibnamefont
  {Harutyunyan}}\ and\ \bibinfo {author} {\bibfnamefont {G.~W.}\ \bibnamefont
  {Milton}},\ }\href {\doibase http://dx.doi.org/10.1007/s00205-016-1034-7}
  {\bibfield  {journal} {\bibinfo  {journal} {Archive for Rational Mechanics
  and Analysis}\ }\textbf {\bibinfo {volume} {223}},\ \bibinfo {pages} {199}
  (\bibinfo {year} {2017}{\natexlab{a}})}\BibitemShut {NoStop}%
\bibitem [{\citenamefont {Harutyunyan}\ and\ \citenamefont
  {Milton}(2017{\natexlab{b}})}]{Harutyunyan:2016:TCE}%
  \BibitemOpen
  \bibfield  {author} {\bibinfo {author} {\bibfnamefont {D.}~\bibnamefont
  {Harutyunyan}}\ and\ \bibinfo {author} {\bibfnamefont {G.~W.}\ \bibnamefont
  {Milton}},\ }\href {\doibase http://dx.doi.org/10.1002/cpa.21699} {\bibfield
  {journal} {\bibinfo  {journal} {Communications on Pure and Applied
  Mathematics (New York)}\ }\textbf {\bibinfo {volume} {70}},\ \bibinfo {pages}
  {2164} (\bibinfo {year} {2017}{\natexlab{b}})}\BibitemShut {NoStop}%
\bibitem [{\citenamefont {John}(1987)}]{John:1987:SLP}%
  \BibitemOpen
  \bibfield  {author} {\bibinfo {author} {\bibfnamefont {S.}~\bibnamefont
  {John}},\ }\href {\doibase http://dx.doi.org/10.1103/PhysRevLett.58.2486}
  {\bibfield  {journal} {\bibinfo  {journal} {Physical Review Letters}\
  }\textbf {\bibinfo {volume} {58}},\ \bibinfo {pages} {2486} (\bibinfo {year}
  {1987})}\BibitemShut {NoStop}%
\bibitem [{\citenamefont {Yablonovitch}(1987)}]{Yablonovitch:1987:ISE}%
  \BibitemOpen
  \bibfield  {author} {\bibinfo {author} {\bibfnamefont {E.}~\bibnamefont
  {Yablonovitch}},\ }\href {\doibase
  http://dx.doi.org/10.1103/PhysRevLett.58.2059} {\bibfield  {journal}
  {\bibinfo  {journal} {Physical Review Letters}\ }\textbf {\bibinfo {volume}
  {58}},\ \bibinfo {pages} {2059} (\bibinfo {year} {1987})}\BibitemShut
  {NoStop}%
\bibitem [{\citenamefont {Figotin}\ and\ \citenamefont
  {Kuchment}(1996{\natexlab{a}})}]{Figotin:1996:BGSI}%
  \BibitemOpen
  \bibfield  {author} {\bibinfo {author} {\bibfnamefont {A.}~\bibnamefont
  {Figotin}}\ and\ \bibinfo {author} {\bibfnamefont {P.}~\bibnamefont
  {Kuchment}},\ }\href {\doibase http://dx.doi.org/10.1137/S0036139994263859}
  {\bibfield  {journal} {\bibinfo  {journal} {SIAM Journal on Applied
  Mathematics}\ }\textbf {\bibinfo {volume} {56}},\ \bibinfo {pages} {68}
  (\bibinfo {year} {1996}{\natexlab{a}})}\BibitemShut {NoStop}%
\bibitem [{\citenamefont {Figotin}\ and\ \citenamefont
  {Kuchment}(1996{\natexlab{b}})}]{Figotin:1996:BGSII}%
  \BibitemOpen
  \bibfield  {author} {\bibinfo {author} {\bibfnamefont {A.}~\bibnamefont
  {Figotin}}\ and\ \bibinfo {author} {\bibfnamefont {P.}~\bibnamefont
  {Kuchment}},\ }\href {\doibase http://dx.doi.org/10.1137/S0036139995285236}
  {\bibfield  {journal} {\bibinfo  {journal} {SIAM Journal on Applied
  Mathematics}\ }\textbf {\bibinfo {volume} {56}},\ \bibinfo {pages} {1561}
  (\bibinfo {year} {1996}{\natexlab{b}})}\BibitemShut {NoStop}%
\bibitem [{\citenamefont {Joannopoulos}\ \emph {et~al.}(2008)\citenamefont
  {Joannopoulos}, \citenamefont {Johnson}, \citenamefont {Winn},\ and\
  \citenamefont {Meade}}]{Joannopoulos:2008:ISS}%
  \BibitemOpen
  \bibfield  {author} {\bibinfo {author} {\bibfnamefont {J.~D.}\ \bibnamefont
  {Joannopoulos}}, \bibinfo {author} {\bibfnamefont {S.~G.}\ \bibnamefont
  {Johnson}}, \bibinfo {author} {\bibfnamefont {J.~N.}\ \bibnamefont {Winn}}, \
  and\ \bibinfo {author} {\bibfnamefont {R.~D.}\ \bibnamefont {Meade}},\
  }\href@noop {} {\emph {\bibinfo {title} {Photonic Crystals: Molding the Flow
  of Light}}},\ \bibinfo {edition} {2nd}\ ed.\ (\bibinfo  {publisher}
  {Princeton University Press},\ \bibinfo {address} {Princeton, New Jersey},\
  \bibinfo {year} {2008})\ pp.\ \bibinfo {pages} {xviii + 286}\BibitemShut
  {NoStop}%
\bibitem [{\citenamefont {Rechtsman}\ and\ \citenamefont
  {Torquato}(2009)}]{Rechtsman:2009:MOU}%
  \BibitemOpen
  \bibfield  {author} {\bibinfo {author} {\bibfnamefont {M.~C.}\ \bibnamefont
  {Rechtsman}}\ and\ \bibinfo {author} {\bibfnamefont {S.}~\bibnamefont
  {Torquato}},\ }\href {\doibase http://dx.doi.org/10.1103/PhysRevB.80.155126}
  {\bibfield  {journal} {\bibinfo  {journal} {Physical Review B: Condensed
  Matter and Materials Physics}\ }\textbf {\bibinfo {volume} {80}},\ \bibinfo
  {pages} {155126} (\bibinfo {year} {2009})}\BibitemShut {NoStop}%
\bibitem [{\citenamefont {Lipton}\ and\ \citenamefont
  {Jr.}(2017)}]{Lipton:2017:BWC}%
  \BibitemOpen
  \bibfield  {author} {\bibinfo {author} {\bibfnamefont {R.}~\bibnamefont
  {Lipton}}\ and\ \bibinfo {author} {\bibfnamefont {R.~V.}\ \bibnamefont
  {Jr.}},\ }\href {\doibase http://dx.doi.org/10.1051/m2an/2016046} {\bibfield
  {journal} {\bibinfo  {journal} {ESAIM: Mathematical Modelling and Numerical
  Analysis}\ }\textbf {\bibinfo {volume} {51}},\ \bibinfo {pages} {889}
  (\bibinfo {year} {2017})}\BibitemShut {NoStop}%
\bibitem [{\citenamefont {Wilcox}(1978)}]{Wilcox:1978:TBW}%
  \BibitemOpen
  \bibfield  {author} {\bibinfo {author} {\bibfnamefont {C.~H.}\ \bibnamefont
  {Wilcox}},\ }\href {\doibase http://dx.doi.org/10.1007/BF02790171} {\bibfield
   {journal} {\bibinfo  {journal} {Journal d'analyse math{\'e}matique}\
  }\textbf {\bibinfo {volume} {33}},\ \bibinfo {pages} {146} (\bibinfo {year}
  {1978})}\BibitemShut {NoStop}%
\bibitem [{\citenamefont {Kuchment}(1993)}]{Kuchment:1993:FTP}%
  \BibitemOpen
  \bibfield  {author} {\bibinfo {author} {\bibfnamefont {P.}~\bibnamefont
  {Kuchment}},\ }\href@noop {} {\emph {\bibinfo {title} {Floquet Theory for
  Partial Differential Equations}}},\ \bibinfo {series} {Operator Theory,
  Advances and Applications}, Vol.~\bibinfo {volume} {60}\ (\bibinfo
  {publisher} {Birkh{\"a}user Verlag},\ \bibinfo {address} {Basel,
  Switzerland},\ \bibinfo {year} {1993})\ p.\ \bibinfo {pages}
  {354}\BibitemShut {NoStop}%
\bibitem [{COM(2013)}]{COMSOL:2013:AMU}%
  \BibitemOpen
  \href
  {http://hpc.mtech.edu/comsol/pdf/Acoustics_Module/AcousticsModuleUsersGuide.pdf}
  {\emph {\bibinfo {title} {Acoustics Module User's Guide, Version 4.3b}}},\
  \bibinfo {organization} {COMSOL},\ \bibinfo {address} {Stockholm, Sweden}
  (\bibinfo {year} {2013}),\ \bibinfo {note} {part number CM020201. See also
  \url{http://www.comsol.com/blogs/theory-thermoacoustics-acoustics-thermal-viscous-losses/}.}\BibitemShut
  {Stop}%
\bibitem [{\citenamefont {Pierce}(1981)}]{Pierce:1981:AIP}%
  \BibitemOpen
  \bibfield  {author} {\bibinfo {author} {\bibfnamefont {A.~D.}\ \bibnamefont
  {Pierce}},\ }\enquote {\bibinfo {title} {Acoustics: An introduction to its
  physical principles and applications},}\ \ (\bibinfo  {publisher}
  {Mc{\-}Graw-Hill},\ \bibinfo {address} {New York, NY, USA},\ \bibinfo {year}
  {1981})\ pp.\ \bibinfo {pages} {515--517}\BibitemShut {NoStop}%
\bibitem [{\citenamefont {Dukhin}\ and\ \citenamefont
  {Goetz}(2009)}]{Dukhin:2009:BVC}%
  \BibitemOpen
  \bibfield  {author} {\bibinfo {author} {\bibfnamefont {A.~S.}\ \bibnamefont
  {Dukhin}}\ and\ \bibinfo {author} {\bibfnamefont {P.~J.}\ \bibnamefont
  {Goetz}},\ }\href {\doibase http://dx.doi.org/10.1063/1.3095471} {\bibfield
  {journal} {\bibinfo  {journal} {Journal of Chemical Physics}\ }\textbf
  {\bibinfo {volume} {130}},\ \bibinfo {pages} {124519} (\bibinfo {year}
  {2009})}\BibitemShut {NoStop}%
\bibitem [{\citenamefont
  {Chandrasekharaiah}(1986)}]{Chandrasekharaiah:1986:ACT}%
  \BibitemOpen
  \bibfield  {author} {\bibinfo {author} {\bibfnamefont {D.~S.}\ \bibnamefont
  {Chandrasekharaiah}},\ }\href {\doibase http://dx.doi.org/10.1115/1.3143705}
  {\bibfield  {journal} {\bibinfo  {journal} {Applied Mechanics Reviews}\
  }\textbf {\bibinfo {volume} {39}},\ \bibinfo {pages} {355} (\bibinfo {year}
  {1986})}\BibitemShut {NoStop}%
\bibitem [{\citenamefont {Norris}(1994)}]{Norris:1994:DGF}%
  \BibitemOpen
  \bibfield  {author} {\bibinfo {author} {\bibfnamefont {A.~N.}\ \bibnamefont
  {Norris}},\ }\href {\doibase http://dx.doi.org/10.1098/rspa.1994.0134}
  {\bibfield  {journal} {\bibinfo  {journal} {Proceedings of the Royal Society
  of London. Series A, Mathematical and Physical Sciences}\ }\textbf {\bibinfo
  {volume} {447}},\ \bibinfo {pages} {175} (\bibinfo {year}
  {1994})}\BibitemShut {NoStop}%
\bibitem [{\citenamefont {Mindlin}(1951)}]{Mindlin:1951:IRI}%
  \BibitemOpen
  \bibfield  {author} {\bibinfo {author} {\bibfnamefont {R.~D.}\ \bibnamefont
  {Mindlin}},\ }\href {http://en.journals.sid.ir/ViewPaper.aspx?ID=294064}
  {\bibfield  {journal} {\bibinfo  {journal} {Journal of Applied Mechanics}\
  }\textbf {\bibinfo {volume} {18}},\ \bibinfo {pages} {31} (\bibinfo {year}
  {1951})}\BibitemShut {NoStop}%
\bibitem [{\citenamefont {Larsen}\ \emph {et~al.}(2009)\citenamefont {Larsen},
  \citenamefont {Laksafoss}, \citenamefont {Jensen},\ and\ \citenamefont
  {Sigmund}}]{Larsen:2009:TML}%
  \BibitemOpen
  \bibfield  {author} {\bibinfo {author} {\bibfnamefont {A.~A.}\ \bibnamefont
  {Larsen}}, \bibinfo {author} {\bibfnamefont {B.}~\bibnamefont {Laksafoss}},
  \bibinfo {author} {\bibfnamefont {J.~S.}\ \bibnamefont {Jensen}}, \ and\
  \bibinfo {author} {\bibfnamefont {O.}~\bibnamefont {Sigmund}},\ }\href
  {\doibase http://dx.doi.org/10.1007/s00158-008-0257-0} {\bibfield  {journal}
  {\bibinfo  {journal} {Structural and Multidisciplinary Optimization}\
  }\textbf {\bibinfo {volume} {37}},\ \bibinfo {pages} {585} (\bibinfo {year}
  {2009})}\BibitemShut {NoStop}%
\end{thebibliography}
\end{document}